# HEDP: A Method for Early Forecasting Software Defects based on Human Error Mechanisms

**Fuqun Huang**
University of Coimbra
Coimbra, 3030-790, Portugal
huangfuqun@dei.uc.pt

**Lorenzo Strigini**
City University of London
London, EC1V 0HB, UK
lorenzo.strigini.1@city.ac.uk

**Abstract**— As the primary cause of software defects, human error is the key to understanding, and perhaps to predicting and avoiding them. Little research has been done to predict defects on the basis of the cognitive errors that cause them. This paper proposes an approach to predicting software defects, so that they may be more easily avoided and/or removed, through knowledge about the cognitive mechanisms of human errors. Our theory is that the main process behind a software defect is that an *error-prone scenario* triggers *human error modes*, which psychologists have observed to recur across diverse activities. Software defects can then be predicted by identifying such scenarios, guided by this knowledge of typical error modes. Compared to current "defect prediction models" that provide a relative likelihood that a program module may contain defects, the proposed idea emphasizes predicting the exact location and form of a possible defect. We conducted two case studies to demonstrate and validate this approach, with 55 programmers in a programming competition and 5 analysts serving as the users of the approach. We found it impressive that the approach was able to predict, at the requirement phase, the exact locations and forms of 7 out of the 22 (31.8%) specific types of defects that were found in the code. The defects predicted tended to be common defects: their occurrences constituted 75.7% of the total number of defects in the 55 developed programs; each of them was introduced by at least two persons. The fraction of the defects introduced by a programmer that were predicted was on average (over all programmers) 75%. Furthermore, these predicted defects were highly persistent through the debugging process. If the prediction had been used to successfully prevent these defects, this could have saved 46.2% of the debugging iterations. This excellent capability of forecasting the exact locations and forms of possible defects at the early phases of software development recommends the approach for substantial benefits to defect prevention and early detection.

## 1  INTRODUCTION

Software defect prediction plays a significant role in software engineering. For instance, it can be used to help allocate testing resources more effectively, focusing on those parts of the software that are predicted as likely to contain faults. The name "prediction" might suggest that we could expect more: that if we can accurately predict an adverse event (the creation of a defect), we can then also prevent it from happening. However, such expectation has not yet met by the existing software defect prediction approaches, though many have been proposed over the last 45 years (N. Fenton et al., 2007; N. E. Fenton & Neil, 1999; Nguyen, Nguyen, & Phuong, 2011; Yang, Tang, & Yao, 2015).

The predictors used in the existing defect predictions models can be categorized into three groups (N. E. Fenton & Neil, 1999): program metrics such as program size and complexity, testing metrics, and software development process metrics. These predictors are then related to defect density by various methods, which have been evolving from simple correlation functions (B. A. Kitchenham, Pickard, & Linkman, 1990) to multivariate approaches such as regression analysis (Nagappan & Ball, 2005), data mining (Menzies, Greenwald, & Frank, 2007) and machine learning algorithms (Elish & Elish, 2008). Graphic methods such as Bayesian Belief Networks (BBNs) (N. Fenton et al., 2007; N. E. Fenton & Neil, 1999) and dependency graphs (Zimmermann & Nagappan, 2008) have been used to analyze the dependencies between various metrics. The contributions of software functionality (Nguyen et al., 2011), project contexts (Rahman, Posnett, & Devanbu, 2012) and code change (Herzig, Just, & Zeller, 2016) have also been examined in recent years. Despite the significant progress made, the relations in these models describe correlation rather than causality. The causal mechanisms underlying software defects are not examined in detail: so, the current defect prediction models can only provide outputs like "module A is likely to contain defects"; they are unable to predict the exact location and form of a likely software defect, e.g., in the form of "requirement A is likely to trigger programmers to introduce a defect B", which could allow focused and effective preventive action.

As the primary cause of software defects, human error is arguably the key to predicting and preventing software defects. Programming is a special type of writing, performed by programmers (Weinberg, 1971); a computer program is a pure cognitive product that describes its designers' thoughts (Détienne, 2002; F. Huang, Liu, & Huang, 2012; F. Huang, Liu, & Wang, 2013). Software defects are the manifestations of cognitive errors of individual software practitioners or/and of miscommunication between software practitioners (Détienne, 2002; F. Huang et al., 2012). However, there is a lack of theory on how software defects are introduced by cognitive error mechanisms and how we can use this theory to predict software defects.

This paper proposes to predict software defects through underlying cognitive error mechanisms. The approach "**H**uman-**E**rror-based **D**efect **P**rediction" (**HEDP**) is proposed to forecast the conditions under which a programming error tends to occur, and the particular form that the error will take. The rest of this paper is organized as follows: Section II reviews the related work on software defect prediction; Section III presents the proposed approach; Section IV presents a case study; Section V discusses the implications, limitations and future studies; Section VI presents the conclusion.

## 2  RELATED WORK

We have found no literature on approaches similar to that studied here, that is, aiming at pinpointing where and in which form defects are likely in a program; we briefly outline the huge literature about software defect prediction in the sense of "identification of subsets of software that are most likely to contain defects".

The software defect prediction area has been continuously developing since the early 1970s, when Akiyama first built a correlation model between lines of code (LOC) and number of defects contained in program modules. More complex metrics such as Cyclomatic complexity (McCabe, 1976) and Halstead complexity (Halstead, 1977) were then proposed to represent the complexity of a software system or module for such correlation models. Since then, both the predictors and the methods used to model the relation between predictors and software defects have evolved.

Various metrics have been used as predictors for defect prediction, such as process metrics (Madeyski & Jureczko, 2015), testing metrics, design metrics (El Emam, Melo, & Machado, 2001), organizational structure metrics (Nagappan, Murphy, & Basili, 2008), code change metrics (Herzig et al., 2016; Moser, Pedrycz, & Succi, 2008), dependency metrics (Zimmermann & Nagappan, 2008) and social network measures (Zanetti, Scholtes, Tessone, & Schweitzer, 2013). With so many metrics, the problem arises: how to choose the most effective metrics? A series of studies try to rank or simplify the metrics using statistical analysis (He, Li, Liu, Chen, & Ma, 2015), while others try to integrate various metrics (Peng, Kou, Wang, Wu, & Shi, 2011), or summarize and compare various metrics through literature review (B. Kitchenham, 2010; Radjenović, Heričko, Torkar, & Živkovič, 2013). A variety of modeling methods have also been proposed to relate the predictors and software defects, such as Bayesian Network (N. E. Fenton & Neil, 1999; G. J. Pai & Dugan, 2007; G. J. Pai, Dugan, J. B, 2007), and various machine learning algorithms (Elish & Elish, 2008; Laradji, Alshayeb, & Ghouti, 2015; Liu, Miao, & Zhang, 2014; Lu, Kocaguneli, & Cukic, 2014) .

Since software prediction should be conducted as early as possible to aid software quality assurance plans (N. Fenton et al., 2008), researchers proposed to predict software defects in various phases of software development (Bhattacharya, Iliofotou, Neamtiu, & Faloutsos, 2012; Ekanayake, Tappolet, Gall, & Bernstein, 2012; N. Fenton et al., 2007). These methods generally use Bayesian networks (N. Fenton et al., 2008) and other graphical models (Bhattacharya et al., 2012) to integrate different kinds of metric data. The most difficult problem in current software defect prediction studies is that the current methods do not work well on novel projects or projects for which historical data for comparable projects are lacking (Zimmermann, Nagappan, Gall, Giger, & Murphy, 2009). To explore these issues, various cross-project defect prediction models are proposed (Rahman et al., 2012; Zimmermann et al., 2009); these studies basically propose to consider more contextual factors (such as whether the products are from the same domain and from the same company) to characterize software projects.

More recently, the importance of software developers' contribution to software defects has been recognized (Bell, Ostrand, & Weyuker, 2013; Weyuker, Ostrand, & Bell, 2008), the number of low-expertise developers (Bird, Nagappan, Murphy, Gall, & Devanbu, 2011) and developer-module networks (Pinzger, Nagappan, & Murphy, 2008) are then proposed to enhance defect prediction models. There is still debate whether the metrics used for human factors are too simplistic to correlate well with defect tendency (Bell et al., 2013; Matsumoto, Kamei, Monden, Matsumoto, & Nakamura, 2010; Pinzger et al., 2008).

Despite the significant progress made, a fundamental limit of the defect prediction methods remains: they can only provide a level of bug-proneness for a unit of source code (i.e., classify a module into a binary variable "defect-prone or not", ranking modules, or the number of defects that a module may contain), but cannot forecast, in the early phases before software implementation, what specific bugs a program is likely to contain. This is because the current prediction models do not exploit any understanding of the causal mechanisms underlying software defects; they are in nature correlation models rather than causal models. Though various procedures, such as regression analysis and machine learning algorithms, are used to model the relationship between predictors and software defects, the basic relations identified are still correlational in nature. As a result, the existing prediction models can only be used to aid in prioritizing test case (P. L. Li, Herbsleb, Shaw, & Robinson, 2006; Nam, 2014; Yoo & Harman, 2012) and planning maintenance (P. Li, Shaw, & Herbsleb, 2003), not to *prevent* the introduction of software defects. Software defect prediction is reported to have very few uses in industrial projects (Engström, Runeson, & Wikstrand, 2010; Lewis et al., 2013; Nam, 2014), and not to yield benefits for developers, due to the lack of "actionable" advice from predictions (Lewis et al., 2013; Nam, 2014). Instead, the new method proposed in this paper promises to (1) support preventing the introduction of defects, by highlighting specification parts likely to cause error, or (2) allow the focused specification of test cases to remove those specific defects, if they occur.

At the same time, various authors have explored the relevance of a developer's characteristics (such as cognitive styles) to the number of errors the developer commits (F. Huang, Liu, Song, & Keyal, 2014; Sharp, Hall, & Bowes, 2015; Westerman, Shryane, Crawshaw, Hockey, & Wyatt-Millington, 1995), or produced taxonomies of errors for aiding conventional defect prevention (F. Huang et al., 2012) or requirement review processes (Anu, Walia, Hu, Carver, & Bradshaw, 2016). A recent interesting study (Castelhano et al., 2018) finds that the insula--a region in the cerebral cortex of the brain-- is highly active when one's error detection process is activated. However, these studies still lack a causal mechanism model explaining how various errors lead to specific software defects in different contexts. As a result, no existing method is capable of forecasting software defects early (before code is produced) with enough precision to allow for focused preventative action.

## 3   THE PROPOSED APPROACH

Though human errors appear to be extremely diverse across different activities, a large proportion of human errors are predictable, in the sense that human errors take a limited number of recognizable patterns (Reason, 1990). The proposed approach HEDP predicts software defects through Error-Prone Scenario Analysis (EPSA), which analyses a programming task for patterns that match the conditions known to trigger erroneous human behaviors; patterns that psychologists have observed to recur across diverse activities (Byrne & Bovair, 1997; Reason, 1990). Section 3.1 defines the concepts used in HEDP; Section 3.2 extracts a set of human error patterns and the general conditions that tend to trigger these error patterns; Section 3.3 presents how to use these error patterns to predict software defects through EPSA.

### 3.1 Terminology

The concepts used in HEDP are defined as follows:

**Defect:** an incorrect or missing step, process, or data definition in a computer program (adapted from (Board, 1990)).

**Error:** an erroneous human behavior that leads to a software defect. Errors are classified at a finer-grained level in psychology as mistakes, slips or lapses (Reason, 1990). Mistakes affect the analysis of a problem or conscious choice of action to perform; slips and lapses are involuntary deviations or omissions in performing the intended action.

**Error Mode (EM):** a particular pattern of erroneous behavior that recurs across different activities, due to the cognitive weakness shared by all humans, e.g. applying "strong-but-now-wrong" rules (see Table 1) (Reason, 1990).

**Error-Prone Scenario (EPS)**: A set of conditions under which an EM tends to occur. An EPS encompasses not only the exterior conditions surrounding an individual (e.g. the most important factor--task), but also the interior cognitive conditions relevant to individual's performance, e.g. his/her knowledge relevant to the task.

**Error-Prone Scenario Analysis (EPSA)**: the activity of an analyst identifying Error-Prone Scenarios.

**Error Mechanism:** how an error is formed; the way causal factors (e.g. the scenario and the error mode) interact to form an error. **The mechanism underlying a software defect is that the *Error-Prone Scenarios* have *triggered* one or more *Error Modes*.** Some defects can be caused by one single error mode, while others may be introduced by a combination of several error modes.

In summary, "Error Mode" concerns "*why*" a defect is introduced; "Error-Prone Scenario" concerns "*when*" (under what circumstances) a defect is introduced; "Error Mechanism" integrates all the aspects concerning "*how*" a defect is introduced.

Several psychological concepts will also be used in the paper:

**Rasmussen's performance level** (Rasmussen, 1983): A framework which classifies cognitive activities into three levels: Skill-based (SB) level, Rule-based (RB) and Knowledge-based (KB) level. We recall the three definitions below. Different performance levels have different cognitive characteristics, thus have different error modes (Reason, 1990).

**Skill-based performance** follows from the statement of an intention, "rolls along" automatically without conscious control. Skill-based activities in programming include typing a text string, compiling a program by pressing a button in the programming environment.

**Rule-based performance** is applicable for tackling familiar problems. It is typically controlled by stored rules or procedures that have been derived from a person's experiences. The mind matches patterns in the situation at hand to the preconditions for such stored rules, allowing quick selection of actions. In programming, there are many rule-based performances, such as programming the printing of a string line, and defining a variable in one's familiar programming language.

**Knowledge-based performance** comes into play when one faces novel situations, and no rules are available from previous experiences. At this level, actions must be planned using analytical process. Errors at this level arise from resource limitations and incomplete or incorrect knowledge. In programming, cognitive performances such as constructing the mental model of the system so as to understand a specific programming task, and trying to figure out a solution for a novel problem are knowledge-based performance.

**Schema**: "A schema is conceived of as a modifiable information structure that represents generic concepts stored in memory. Schemata represent knowledge that we experience--interrelationships between objects, situations, events, and sequences of events that normally occur. In this sense, schemata are prototypes in memory of frequently experienced situations that individuals use to interpret instances of related knowledge." (Glaser, 1984)

### 3.2 Human Error Mechanisms

A model of the Human Error Mechanism describing the process of how a software defect is caused by human errors is proposed in Fig. 1. This model includes the main causal factors that determine a human error (Reason, 1990): the nature of task (including the representation forms and the content of a task), the nature of the programmer (mainly including their current available knowledge base (F. Huang et al., 2013)), and human error modes, which are the general mechanisms governing humans' erroneous cognitive performance.

A software defect is caused when the conditions of the programming task and/or programmer trigger one or more human error modes. To predict an error, the task and individual should be analyzed together because these two factors interact. For instance, the same task could be very easy for one person while difficult for another person. This integrated feature of a task and a human individual is modeled by a "scenario".

Based on the human error mechanism model in Fig. 1, software defects can then be predicted through identifying in the current programming contexts (e.g., a program specification) conditions that are likely to trigger human error modes. To conduct such prediction, the first step is to build a pool of human error mechanisms that contains the error modes and their associated general scenarios.

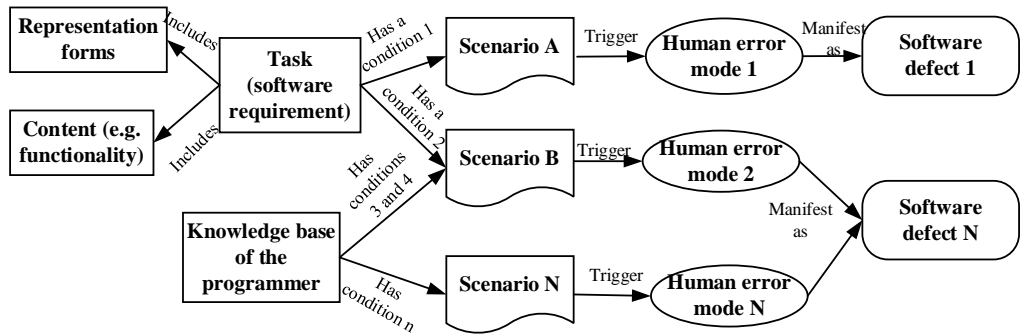

Fig. 1. A model of human error mechanisms underlying software defects, with examples of possible relationships between

We developed a pool of human error mechanisms (shown in Table 1) based on the error patterns that have been widely accepted in the psychological community, such as the error patterns summarized by J. Reason (Reason, 1990) and Byrne and Bovair's theory on "post-completion errors"(Byrne & Bovair, 1997). These error modes are included because they are reported to reoccur across diverse activities. Three strategies are used to build the pool of human error mechanisms:

1). Error modes are only included if they are suitable for EPSA. Some of the error modes identified in psychology are not. For example, when an individual's working memory is overloaded, he/she is likely to commit an error; this is called "Workspace Limitations" by Reason (Reason, 1990). However, working memory load is closely related to one's working memory capacity, one's expertise on the problem (intrinsic load), the format of the problem representation materials (extraneous load) and one's schema construction processes (germane load) (Schnotz & Kurschner, 2007). Working memory overload is a real-time cognitive state that is thus hard to forecast. Therefore, we do not include it in our initial pool of human error mechanisms; adding it may become possible with further study of cognitive load in programming tasks.

2). The error mode taxonomy is adapted in order to retain a systematic relation between the error modes included. For example "Countersigns and non-signs", "Rule strength", and "General rules", and "Redundancy" are four sub-modes under the mode "misapplication of good rules" in Reason's work (Reason, 1990). These modes describe different situations in which people are prone to apply "strong-but-now-wrong" rules, but they pertain to a same mechanism. Therefore, we consider them not as error modes but as scenarios in which the error mode of "applying 'strong-but-now-wrong' rule" is prone to occur.

3). The Error Modes are extracted and represented by pseudo codes, shown in the third column of Table 1. The fundamental psychological theories tend to be somewhat vague and thus difficult to apply for practical purposes. We specified the preconditions (using "IF"), the situations under which an error tends to occur (using "WHEN"), and the final manifestation of the error (using "THEN"). Notations such as "AND" and "OR" are used to combine multiple situations. The definitions of the specific notations other than natural language are provided in Table 2.

TABLE 1

A POOL OF HUMAN ERROR MECHANISMS

| Error modes | Descriptions of the error modes | Scenario representations | |
|---|---|---|---|
| Applying "strong-but-now-wrong" rule | In a context that is similar to past circumstances, people tend to behave in the same way, neglecting the signs of exceptional or novel circumstances. The typical scenarios include: the person encounters a new feature for the first time ("First Exception"); the competing rule is very "strong", in that has been successfully used many times before ("Rule Strength"); and the competing rule is a general rule ("General Rules") (Reason, 1990). It means that programmers tend to prefer rules that have been successfully used in the past. The more frequently and successfully the rule has been used before, the more likely it is to be recalled and used. | **IF** | Current task requires Rule X <Feature FeX>, |
| | | **WHEN** | There Exists Rule A <Feature FeA, Frequency of successful usage FuA>, **AND** There Exists Rule B <Feature FeB, Frequency of successful usage FuB >; $\{FeX \cap FeB\} \supseteq \{FeX \cap FeA\} \neq \emptyset$; $\quad$ FuA >> FuB, $\quad$ **OR** $\quad$ Fu$((FeX \cap FeB)_i)=0$, $\quad$ **OR** $\quad$ FeB $\subset$ FeA; |
| | | **THEN** | The person tends to retrieve Rule A. |
| Rule encoding[1] deficiencies | Features of a particular situation are either not encoded at all or misrepresented in the conditional component of the rule (Reason, 1990). | **IF** | Current task requires Rule X <Feature FeX>; |
| | | **WHEN** | There Exists Rule $\tilde{X}$<Feature Fe$\tilde{X}$ >; $\quad$ FeX $-$ Fe$\tilde{X}$ $\neq \emptyset$ |
| | | **IF** | The person uses Rule $\tilde{X}$; |
| | | **THEN** | The person commits an error of misusing Rule $\tilde{X}$ for the current situation that requires Rule X. |
| Lack of knowledge | Software faults are introduced when one lacks some knowledge, or even does not realize some other knowledge is required. This error mode is liable to appear especially when the problem belongs to an unfamiliar application domain. For example, the international Date Line problem in navigation software systems of F-22 Raptor(Daily, 2007). | **IF** | Current task requires Rule X; |
| | | **WHEN** | Rule X does not exist; |
| | | **THEN** | The person tends to fail the task. |
| "Difficulties with exponential developments" (Reason, 1990) | "Processes that develop exponentially have great significance for systems in either growth or decline, yet subjects appeared to have o intuitive feeling for them. When asked to gauge such processes, they almost invariably underestimated their rate of change and were constantly surprised at their outcomes." This means humans tend to construct linear models (whose growth rate is lower than exponential models) when exponential models are required to understand a situation in reality (Reason, 1990). | **IF** | Current task requires extracting a relation between independent variable x and dependent variable y according to a sample data; |
| | | **WHEN** | The actual relation belongs to models in the families of "$y = x^p$" or "$y = d^x$"; |
| | | **THEN** | People tend to construct wrong models in the family of "$y = ax$". |
| Selectivity | Psychologically salient, rather than logically important task information is attended to (Reason, 1990). "Selectivity" in programming means that when a programmer is solving problems, if attention is given to the wrong features or not given to the right features, mistakes will occur, resulting in wrong problem presentation, or selecting wrong rules or schemata to develop solutions. For example, if there are several requirement items scattered at different places in a document, for a single task, programmer may fail to notice some information and perceive a wrong representation of the task. | **IF** | Current task contains features FeT$_i$ ={Salience$_i$, LogicImportance$_i$} and FeT$_j$ ={Salience$_j$, LogicImportance$_j$} |
| | | **WHEN** | Salience$_j$ > Salience$_i$ ; |
| | | **THEN** | People tend to notice a FeT$_j$; |
| | | **WHEN** | LogicImportance$_i$ > LogicImportance$_j$ ; |
| | | **THEN** | An error tends to be introduced due to the omission of FeT$_i$. |
| Biased review | Humans tend to believe that all possible courses of action have been considered, when in fact only a subset have been considered. When programmers generate test cases in debugging, they may fail to take all conditions into | **IF** | Current task requires a person to review one's own work X; |
| | | **WHEN** | X contains N courses or conditions; |
| | | **THEN** | The person tends to review n<N courses or conditions. |

[1] Encoding is the process of extracting features of a situation and mentally representing the situation based on the extracted features.

| | | | |
|---|---|---|---|
| | consideration, e.g. exception and boundary conditions (Reason, 1990). | | |
| Post-completion error (Byrne & Bovair, 1997) | If the ultimate goal is decomposed into sub-goals, a sub-goal is likely to be omitted under the following conditions: the sub-goal is not a necessary condition for the achievement of its super-ordinate goal, and the sub-goal is to be carried out at the end of the task. One well-known example is that, if the work flow of an Automatic Teller Machine (ATM) is designed as that withdrawing bank card is the last step, people are prone to leave after taking the cash but forget to withdraw their card from the ATM. | IF<br>WHEN<br><br><br><br><br>THEN | Task A ={Task A.1, Task A.2};<br><Task A.1 is the main subtask>,<br>  AND<br><Task A.2 is not a necessary condition to Task A.1>,<br>  AND<br><Task A.2 is the last step of Task A >;<br>Humans tend to omit Task A.2. |

Table 1 is *not* meant to exhaustively enumerate all of the human error modes described in cognitive psychology. This pool is used for the practical purpose of exploring whether and how software defects can be predicted based on human error mechanisms; therefore, it only includes a selected set of human error modes that the authors consider systematic, valid, typical and suitable for performing EPSA, based on the current understanding of human error theories. The pool can be extended with the progress of research in cognitive psychology and with the extension of this study in the future.

TABLE 2

A SAMPLE OF NOTATIONS USED TO REPRESENT HUMAN ERROR SCENARIOS

| Notation | Meaning |
|---|---|
| IF | Starting a clause defining the precondition of an Error-Prone Scenario |
| WHEN | Starting a clause describing the conditions that form an Error-Prone Scenario |
| THEN | Starting a clause describing the manifestation form of an Error |
| AND | Logic AND, collecting two conditions, both of which need to be satisfied for forming an Error-prone Scenario. |
| OR | Logic OR, collecting two conditions, either of which needs to be satisfied for forming an Error-prone Scenario. |
| ∩ | Intersection |
| ⊂ | Belongs to |
| ∅ | Empty set |
| > | More than |
| < | Less than |
| = | Equal |
| ≠ | Not equal |
| >> | Far more than |
| , | Conjunction between conditions that constitute an error-prone scenario |
| ; | Separation between two independent clauses. |
| . | The end of the representation of a human error mode. |

### 3.3 Defect forecasting process

Since the main process behind software defects is that an error-prone scenario triggers human error modes, software defects can be predicted by identifying those current scenarios that tend to trigger the general human error modes. This process of identifying the features of the task and the programmers, and examining whether these features tend to trigger human error modes is EPSA, the key process of HEDP, shown in Fig. 2. HEDP includes three sub-processes:

*Task analysis*. Task analysis aims to answer the questions of "what need to be done" and "how it may be done". This includes reviewing task representation materials (e.g. requirement specification) and the content of the task itself. For a given programming problem, different programmers may generate diverse styles of codes at the detailed level. However, for a given task specification, we conjecture that some parts of the readers' (programmers') mental representation would be shared by many; and for a given task representation, they will typically choose only one, or among few, algorithms to solve it. The task analysis is meant to extract these common task representations and algorithms. An example for task analysis is shown in Section 4.5.1.

*Knowledge assessment*. This is the process to learn about a programmer's schemata (Adelson & Soloway, 1986) stored in his/her knowledge base associated to the task. Accurately assessing an experienced programmer's knowledge is still challenging due to the large volume of knowledge the programmer may possess. For novice student programmers, we may assess their programming knowledge by reviewing the textbooks they used; textbooks should account for a large proportion of novice student programmers' knowledge base. An example of knowledge assessment is shown in Section 4.5.2.

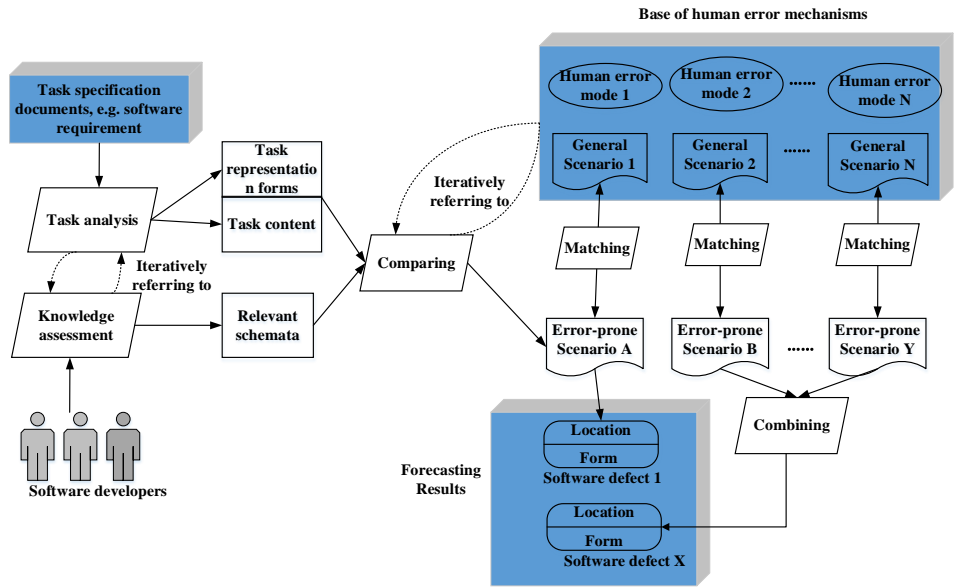

Fig. 2. HEDP framework

*Error-Prone Scenario Analysis.* This is the process to match the specific features of the task and of programmer knowledge to the general features that tend to trigger a human error mode. This can be easy for some error modes that are triggered only by task features, e.g. post-completion error, an error in which humans tend to omit a sub-task that should be carried out at the end of a task but is not a necessary condition for the achievement of the main sub-task (see Table 1). Some error modes require the analysts to know both the task features and the programmer's related knowledge. Considering the error mode "applying 'strong-but-now-wrong' rule" in Table 1, the analyst needs to know what rules will be needed to accomplish a task, and what relevant rules a programmer has.

These three sub-processes are not three sequential steps, but are performed in an iterative pattern. We note that the scenario analysis itself is an analyst's cognitive process. Just as a code inspector needs to have domain knowledge and programming knowledge in order to identify a bug in a code, a HEDP analyst would need to have programming knowledge and the above knowledge of human error mechanisms.

A set of graphic symbols is proposed to guide the scenario analysis and record the results, shown in Fig. 3.

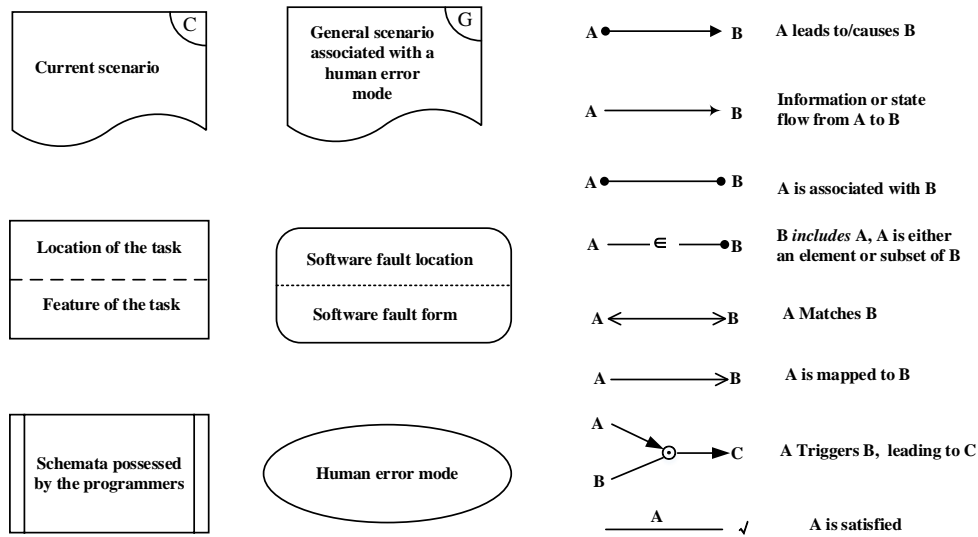

Fig. 3. Symbols used to aid error scenario analysis

Though these symbols are not a necessity for error-prone scenario analysis, we expect such graphical representations to help. Symbols could remind one of the essential elements that constitute an error-prone scenario, so as to promote the thinking process as well as represent the mental model produced during the analyzing process (F. Huang et al., 2013).

## 4 CASE STUDY

An empirical study was conducted (by the first author, Huang) to demonstrate how to use HEDP and evaluate its effectiveness and potential benefit. The empirical study was performed on a programming task, called the "jiong" problem, in a programing contest setting. The study compared the predicted defects using HEDP with the actual defects introduced by programmers.

### 4.1 Overview of the process

The study was conducted in the context of a programming contest as follows:

1) Before the programming contest, the contestants filled in a survey about their programming knowledge background.

2) The contest presented eight programming problems. The contestants could choose to solve up to all of these in four hours. The researchers selected the "jiong" problem to conduct this study, as this problem had been tried by most contestants, and it was used as the task in a concurrent study performed for other purposes (examining the correlations between personality traits, program metrics and number of defects a programmer introduced) (F. Huang et al., 2014). Data reuse is useful here, as it allows interested readers to make comparisons between various models.

3) After the programming contest, the first author predicted defects using HEDP, including Task Analysis performed on the requirement specification, knowledge assessment based on the contestants' knowledge background, and EPSA. *We need to emphasize that EPSA (the prediction) was performed before the researcher accessing to any code.* The process was well documented and reviewed by 5 academic committees in the first author's Ph.D. thesis (F. Huang, 2013), and remains unpublished internal material stored in the library of Beihang University.

4) Concurrently with 3), a software engineer who was independent of this study performed a code inspection to identify the defects introduced by the contestants.

5) Finally, the researchers compared the predicted defects to the actual defects identified by the code inspector.

### 4.2 The context of the study

The study was conducted in the context of the 7th annual programming contest run at Beihang University (BUAA) to select candidates to participate in the Asian Qualifying of the ACM International Collegiate Programming Contest (ACM-ICPC). The contest was held in the form of on-site testing in a computer room. Each contestant was assigned one computer.

There were strict precautions against cheating: 1) each room was monitored by two supervisors to prevent the contestants from copying others' code; 2) the programming environment was cut off from external internet to prevent the contestants from learning or copying code from the internet. With such precautions, and in a competitive contest where the contestants were highly motivated to win, we believe that the contestants worked independently; any similarity between errors by different contestants is unlikely to be due to copying or collaborating.

The contest scores were announced in real time by the Online Judge System, which was similar to the system used in the ACM-ICPC (Meulen & Revilla, 2008). The contestant can first compile and run the program on his/her local environment, then submit it to the Online Judge System on a server. For each problem, contestants can submit to the Online Judge System as many versions of programs as they wish, until the system "accepts" one version or the contestant quits. After each submission, the Online Judge System fed back to the contestants these types of results:

**Accepted (AC)**. The output of the program match what the Online Judge expects.

**Wrong Answer (WA)**. The output of the program does not match what the Online Judge expects.

**Presentation Error (PE)**. The program produces correct output matching the Online Judge's secret data, but does not produce it in the correct format.

**Runtime Error (RE)**. This error indicates that the program performs an illegal operation when running on the Online Judge's input. Some illegal operations include invalid memory references such as access outside an array boundary. There are also a number of common mathematical errors such as "divide by zero" error or "overflow".

**Time Limit Exceeded (TL)**. The Online Judge has a specified time limit for every problem. When the program does not terminate within that time limit, this error will be generated.

**Compile Error (CE)**. The program does not compile with the specified language's compiler.

### 4.3 The Participants

The programming contest was open to all students from Beihang University. The contestants were asked to fill a series of surveys (including the survey relevant to this study--"Survey of Knowledge Background", and other surveys on cognitive styles, relevant to another study in (F. Huang et al., 2014)) voluntarily before the programming contest began, and to give consent for use of their entries for research. The Survey of Knowledge Background included questions about the college grade, major, the courses one has taken, project experiences in programming languages, algorithms, design patterns and programming tools. Our research team sponsored a gift package, which included two energy bars, a bottle of water, papers and a pen. All the contestants were welcomed to take a gift package, no matter whether they were willing to answer the surveys.

The problem chosen by most contestants was the "jiong" problem (shown in Section 4.4). This study concerns the fifty-five contestants among these who answered the surveys and consented to taking part.

According to the responses to the Survey of Knowledge Background, all of the participants were in the first year of an undergraduate course, with majors in computer science or software engineering. All of them had just finished their course on the C language provided by the university, and had no professional programming experience.

### 4.4 The task

The task of this case study is named the "jiong" problem. "Jiong" is a simplified Chinese character (Hanzi) shown in Table 3. The requirement specification the task is shown in Table 3. The line numbers in the left column were not in the original requirements presented to the contestants; they are added by the authors to facilitate the analysis in this paper.

TABLE 3

THE REQUIREMENT SPECIFICATION OF THE "JIONG" TASK

| | |
|---|---|
| *Line1* | Print a Chinese word "jiong" in a nested structure. |
| *L2* | **Inputs** |
| *L3* | There is an integer in the first line that indicates the number of input groups. |
| *L4* | Each input group contains an integer n (1≤n≤7). |
| *L5* | **Outputs** |
| *L6* | Print a word "jiong" after each input group, and then print a blank line after each "jiong" word. |
| *L7* | **Sample Inputs** |
| *L8* | 3 |
| *L9* | 1 |
| *L10* | 2 |
| *L11* | 3 |
| *L12* | **Sample Outputs** |
| *L13* | 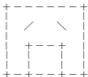 |
| *L14* | |
| *L15* | 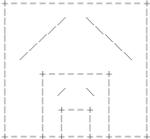 |
| *L16* | |

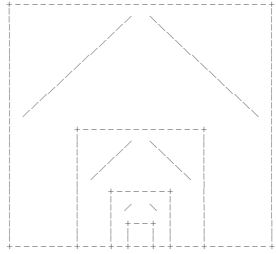

*L17*

*L18*

The test case used by the Online Judge System to assess the "jiong" programs comprised "32" in the first line (as the number of input groups) and numbers in the interval [1,7] (as the nesting level of the "jiong" word in the following 32 lines). This test case achieved 100% requirement coverage for the "jiong" problem. Only when a version had passed the test case and its runtime was equal to or less than 1000 milliseconds would it be accepted by the Online Judge System.

### 4.5 Defect prediction using HEDP

HEDP was performed on the "Jiong" problem's requirement specification to predict possible defects. Details follow in Section 4.5.1, 4.5.2 and 4.5.3.

#### 4.5.1 Task analysis

For a given programming problem, we expect that the programs written by different programmers may differ in detail; but the problem to be solved poses a set of identical demands to all, so that some similar or common parts exist between different programs. For instance, Table 4 shows the two typical solutions of the "jiong" problem and the elements likely to be shared by programs implementing either solution.

TABLE 4
TWO TYPICAL SOLUTIONS OF THE "JIONG" PROBLEM

| | | |
|---|---|---|
| **Solution A** | 1) | Define and initialize an array to store the symbols (+, -, /, \|, blank space) to form the smallest "jiong" as indicated by the image; |
| | 2) | Read in the number of "jiong"s, and the nesting levels for each "jiong" (indicated as n); |
| | 3) | Compute the symbols of the "jiong" at the *n*-th nesting level by recursion according to the relationship revealed in the images, and store in another array; |
| | 4) | Print a "jiong" line by line by two nesting loops, where the number of iterations is related to the width and height of the "jiong". Model the relationship between the structure of the "jiong" and its nesting level: the width=$2^{n+2}$, height = width; |
| | 5) | Print a blank line after the "jiong"; |
| | 6) | Return to the loops and process the next "jiong", until all the "jiong" images are printed. |
| **Solution B** | 1) | Define and initialize an array to store the symbols (+, -, /, \|, blank space) to form the smallest "jiong" as indicated by the image; |
| | 2) | Read in the number of "jiong"s, and the nesting levels for each "jiong" (indicated as n); |
| | 3) | Model the relationship between the structure of the "jiong" and its nesting level: width=$2^{n+2}$, height = width; |
| | 4) | Compute and print each symbol of the "jiong" by iteration, where the type of a symbol is determined by the height of the symbol's location in the "jiong" image; |
| | 5) | Print where the number of iteration is related to the width and height of the jiong"; |
| | 6) | Print a blank line after the "jiong"; |
| | 7) | Return to the loops and process the next "jiong", until all the "jiong" images are printed. |

#### 4.5.2 Knowledge assessment

To perform error-prone scenario analysis, one needs to know the relevant schemata stored in a programmer's knowledge base. Assessing programming knowledge is still a challenging research area (Chatzopoulou & Economides, 2010). It can be very difficult to perform knowledge assessment on experienced programmers, as their knowledge base can be very abundant and diverse. Just reviewing external materials through which the experienced programmers obtained their knowledge will not suffice to completely assess that knowledge. On the other hand, assessing the knowledge of student programmers could be much easier, as their programming experience was mainly gained from university courses, which have external representation materials such as text books. Using student programmers gives us the opportunity to test the philosophy of predicting software defects based on human error mechanisms, which is the focus of this paper. If this philosophy is found to be promising, future studies will be required to address the challenge of detailed knowledge assessment on expert programmers.

Based on the information obtained from the survey of knowledge background, it is reasonable to use the textbook used by these participants' C language course, "C programming" (Tan, 2005), as the external representation of the programming schemata that the participants possessed. All participants were students from the same University and had used the same textbook. The researcher reviewed this textbook and identified the knowledge that may be required for solving the "jiong" problem. Then, error-prone scenario analysis was performed, described in the following section.

### 4.5.3 Error-prone scenario analysis

Based on the task specification in Section 4.1, the knowledge assessment results in Section 4.4.1 and the task analysis results in Section 4.4.2, we performed error-prone scenario analysis. In total, seven error-prone scenarios were identified, with seven scenario forms produced, shown from Table 5 to Table 11. Each form described the location of the scenario within the task, the error modes that were likely to be triggered, the fault forms that were likely to occur, and how this scenario was identified--scenario analysis. Each scenario analysis process produced an error mechanism model, describing possible interaction mechanisms between the task, individuals and error modes.

TABLE 5
ERROR-PRONE SCENARIO ANALYSIS (1)

***Scenario ID:*** ES1

***Defect location:*** The location between the functions of printing one single "jiong" and returning to the loop to process the next "jiong", i.e. Step 5) of Solution A in Table 2.

***Error modes:*** Post-completion error

***Scenario analysis:*** In the process of solving the "jiong" problem, the programmers' ultimate aim (superordinate goal) is to compute and print the image of the "jiong". The task of printing a blank line after the "jiong" is a sub-goal, but it is carried out at the end of the task. This is a typical scenario of post-completion error, where the sub-goal is likely to be omitted, thus, programmers may forget to print the blank line after the "jiong".

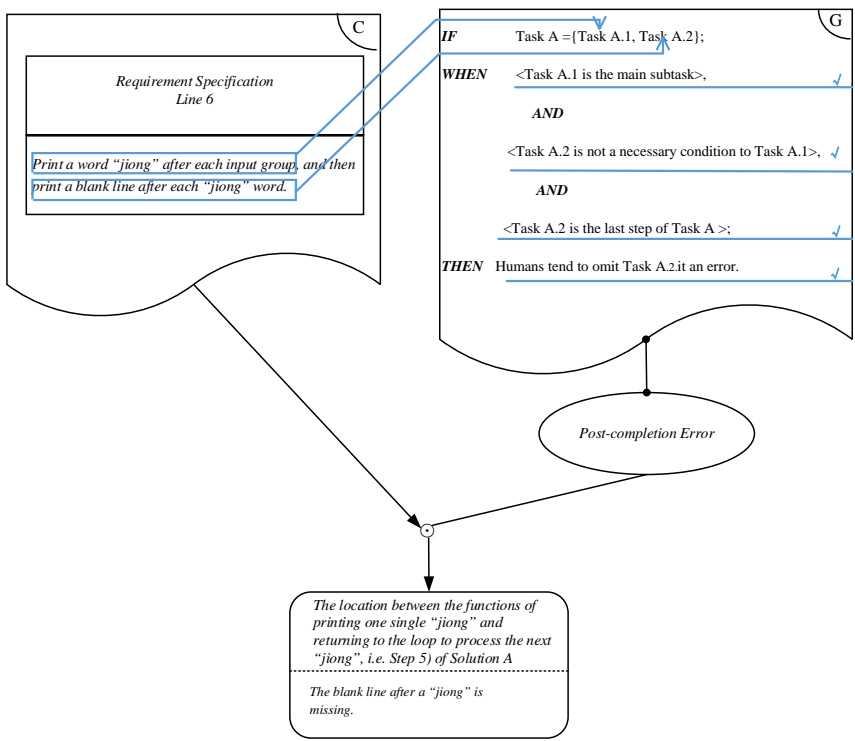

***Defect form:*** the blank line after a "jiong" is missing.

TABLE 6
ERROR-PRONE SCENARIO ANALYSIS (2)

***Scenario ID:*** ES2

***Defect location:*** *The locations where array or variables are defined, e.g. Step 1) of Solution A in Table 2.*

***Error modes:*** *Rule encoding deficiencies*

***Scenario analysis:*** The analyst checked the textbook (Tan, 2005) about the contents of variables and arrays and found that the author represents the contents relevant to variable in several separate sections. The textbook first introduces the structure of a variable very simply with a small image, including *name, value and size.* Then it takes several long sections to describe different *types of variables*. After that, a section about variable assignment is presented, followed by several sections about operators.

No passage in the textbook explicitly represents the full structure of schema for array or variable definition. For the general schema of variable definition, it contains a total of four sub-rules: giving a name to the variable, defining the type of the variable, allocating the variable with a proper size, and initializing the variable. The students need to acquire these sub-rules from different places at different times, and then encode them together to form the complete schema for variable definition. Therefore, the analyst anticipates that some students may have not encoded a few sub-rules or not integrated all the sub-rules to the general rule. If the sub-rule of "initialization" has not been integrated to the general rule of variable definition, the programmer may forget initializing an array or variable when he defines it. Then, a fault is likely to be introduced in the form of "using array or variable without initialization".

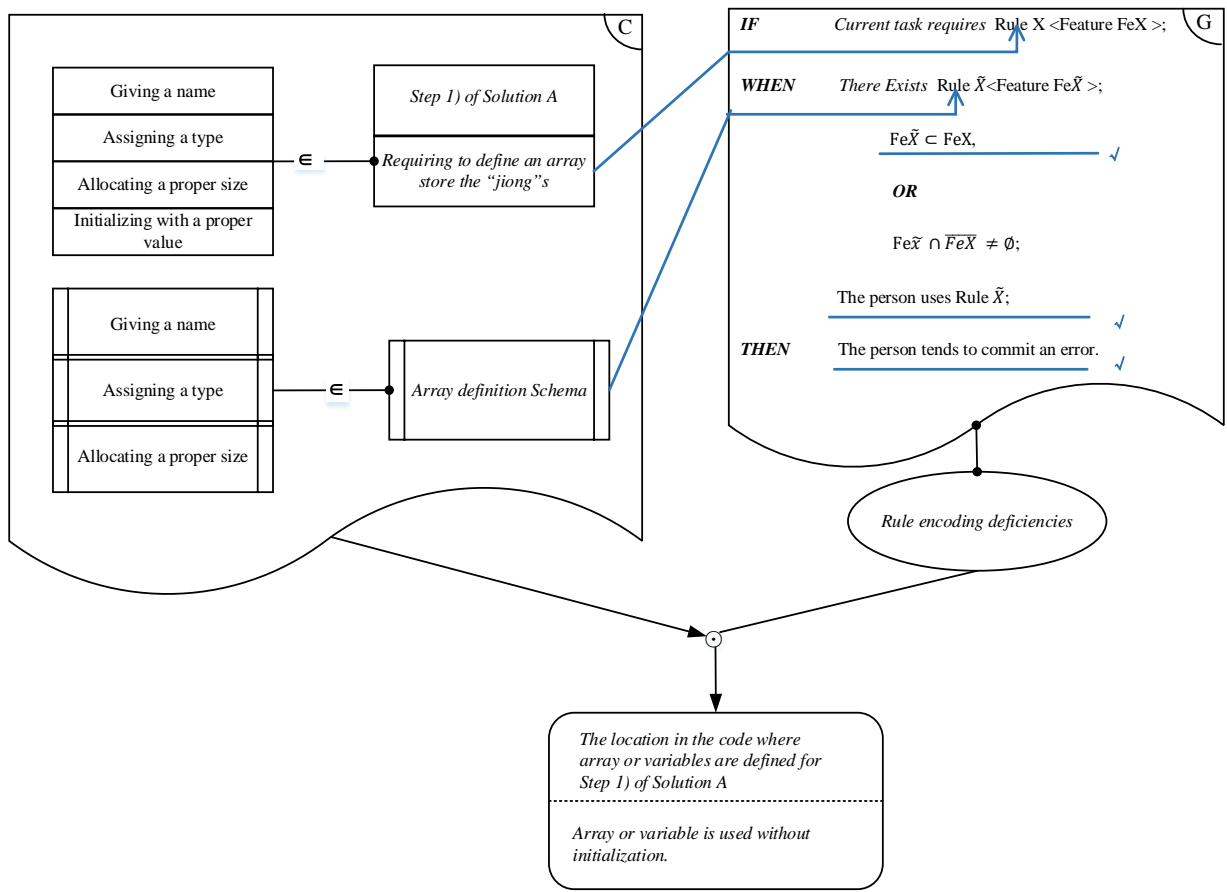

***Defect form:*** *Array or variable is used without initialization.*

TABLE 7

ERROR-PRONE SCENARIO ANALYSIS (3)

***Scenario ID:*** ES3

***Defect location:*** *The locations where arrays are defined, e.g. Step 1) of Solution A in Table 2.*

***Error modes:*** *Rule encoding deficiencies*

***Scenario analysis***: The analyzing process of this scenario is similar to the scenario of ES2. Allocating a variable or array with a proper size is an important sub-rule of the general rule of variable or array definition. The size of the array is determined by the specific application requirement. If the allocated size is too large, it is a waste of memory. If the allocated size is too small, an overflow fault can occur.

Similar as the situations described in ES2: no place in the textbook explicitly presents the complete schema of variable/array definition. The sub-rule may not be encoded at all or not integrated to the general rule for some students. Thus, a fault may occur in the form that the size of array defined is smaller than that is required.

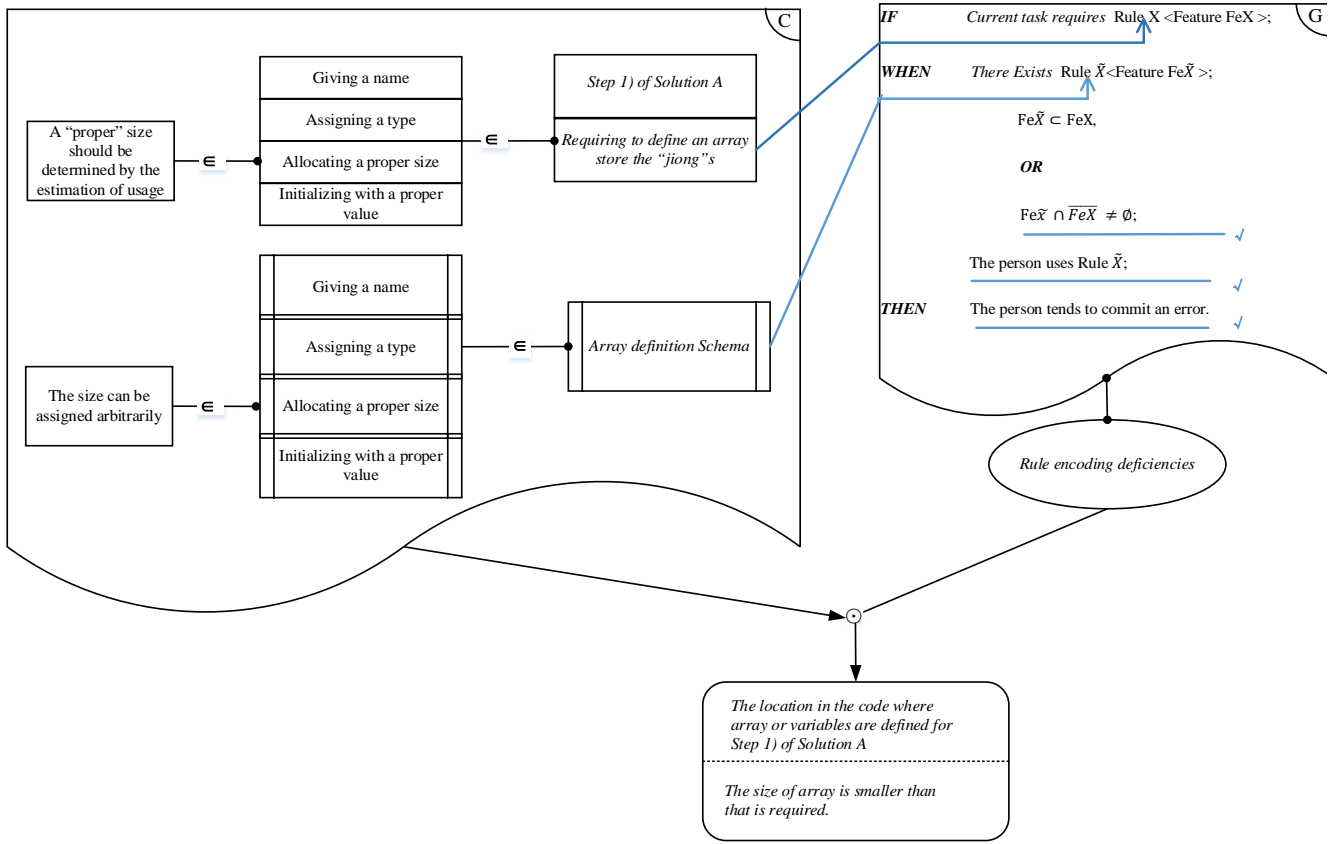

***Defect form:*** The size of array is smaller than that is required.

TABLE 8
ERROR-PRONE SCENARIO ANALYSIS (4)

***Scenario ID:*** ES4

***Defect location:*** *The locations where the array used to store "jiong" is initialized, i.e. Step 1) of Solution A in Table 2.*

***Error modes:*** *Applying "strong-but-now-wrong" rule under the "first exception"*

***Scenario analysis:*** In the array used to store the symbols of the image "jiong", many elements should be assigned blank space. On the other hand, however, the programming examples and exercises in the textbook used by the participants generally require them to initialize array contents with numbers or strings. The students have never experienced initializing a variable to blank spaces before. This is a typical scenario of "first exception". The "first exception" means that on the first occasion an individual encounters a significant exception to a general rule, particularly if that rule has repeatedly shown itself to be reliable in the past, the "strong-but-now-wrong" rule is likely to be applied. Thus, the scenario is prone to trigger the participants applying "strong-but-now-wrong" rules: initializing the array to "0" instead of "blank space".

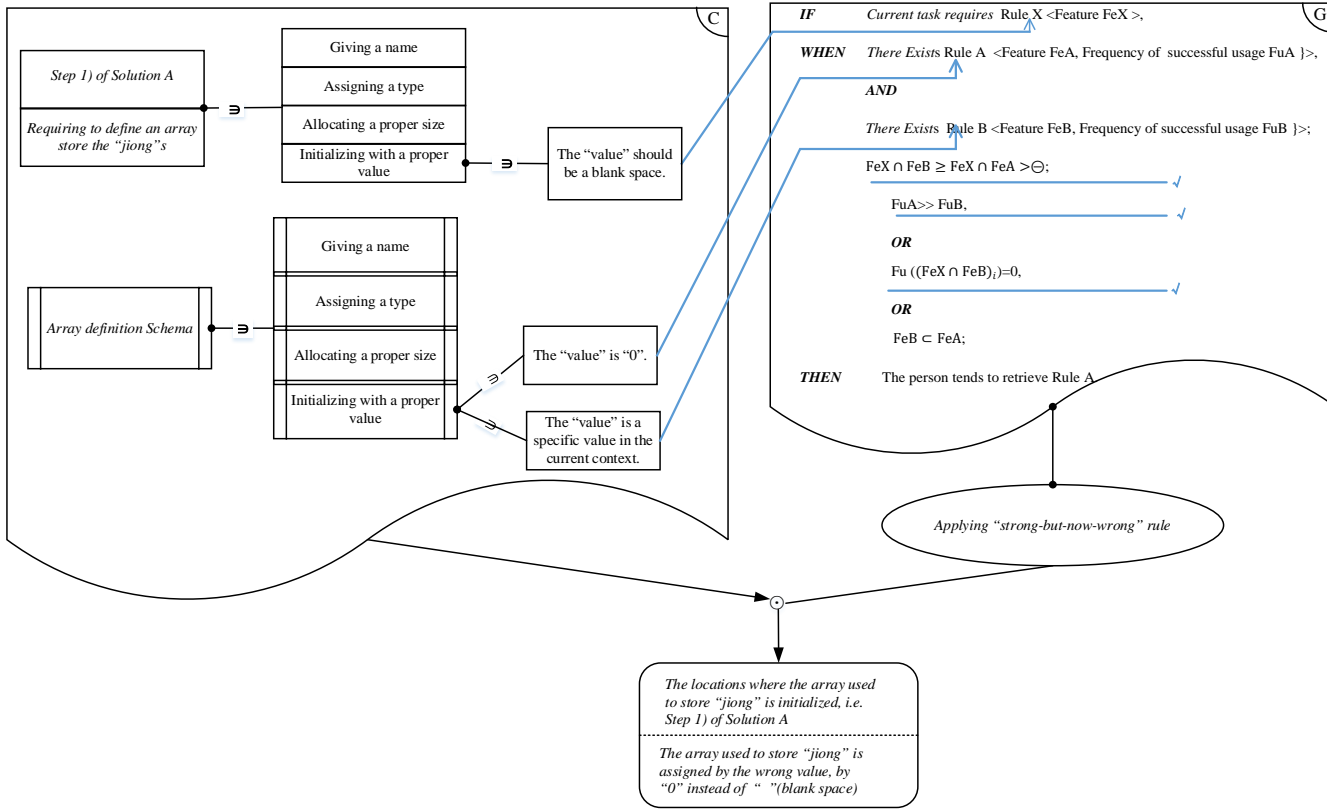

***Defect form:*** *The array used to store "jiong" is initialized by the wrong value, by "0" instead of " " (blank space).*

TABLE 9
ERROR-PRONE SCENARIO ANALYSIS (5)

*Scenario ID:* ES5

*Defect location:* *The locations where the algorithm of calculating "jiong"s is involved, e.g. Step 3), 4) and 6) of Solution A in Table 2.*

*Error modes:* *Lack of knowledge*

*Scenario analysis:* The programming schema of either recursion or iteration is essential to solve the "jiong" problem. If one has not mastered these schemata, one can only enumerate all the "jiong" words (specify and print each character of the arrays used to store the "jiong" words) of different nesting levels. As the nesting level becomes high (e.g.7), the corresponding "jiong" word becomes so large that a person would need more time than was allowed in the contest to produce all this code.

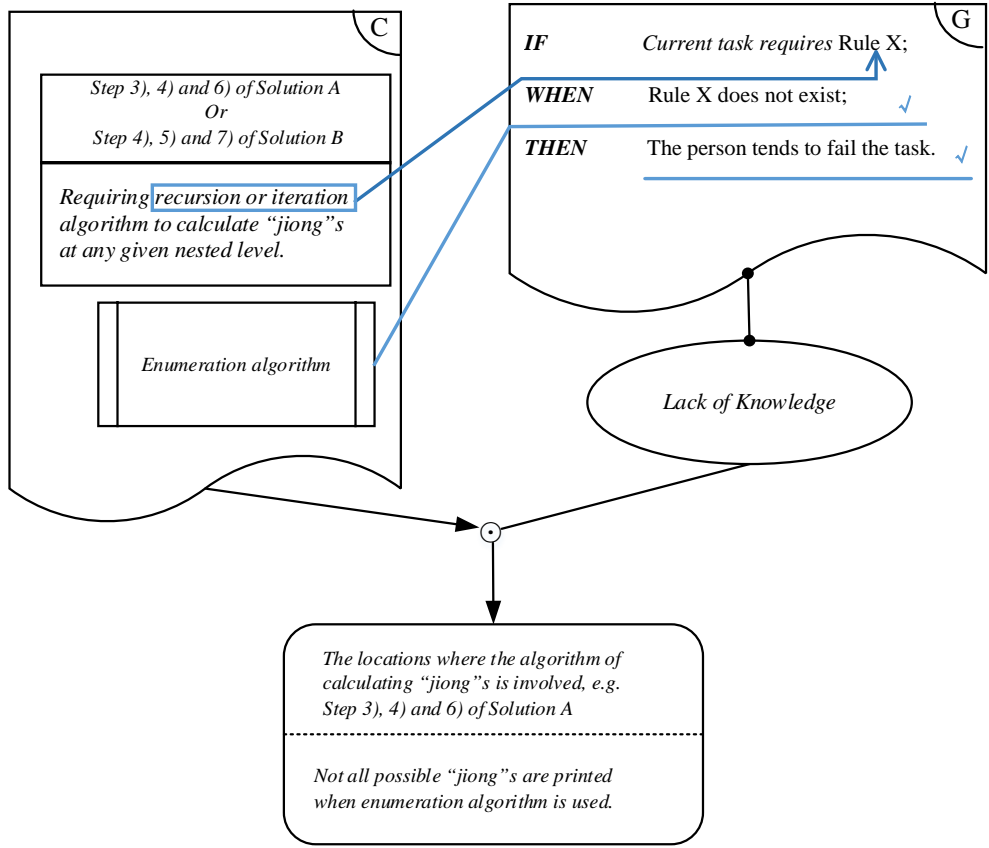

*Defect form:* *Not all possible "jiong"s are printed when enumeration algorithm is used.*

TABLE 10
ERROR-PRONE SCENARIO ANALYSIS (6)

***Scenario ID:*** ES6

***Defect location:*** *The places where the inputs and outputs are addressed, e.g. Step 2) and 6) of Solution A in Table 2.*

***Error modes:*** *Selectivity*

***Scenario analysis:*** There are two passages in the requirement specification that contain information on the formats of inputs and outputs. The first passage is located under the bullets "**Inputs**" and "**Outputs**". Under the bullet "Input", the format of input is specified as "there is an integer in the first line, which indicates the number of input groups. Each input group contains an integer n ($1 \leq n \leq 7$)." Under the bullet "Output", the format of output is specified as "print out a word 'jiong' after each input group, and then print out a blank line."

The other passage relevant to the formats of inputs and outputs is located under "**Sample inputs**" and "**Sample outputs**". Using "Sample inputs" and "Sample outputs" is a conventional part of specification in the programming contest, widely used across different tasks and different annual contests. The "Sample inputs" and "Sample outputs" here are static: they cannot reveal the required sequencing of inputs and outputs. However, if the participants only notice information under the "Sample inputs" and "Sample outputs", they may interpret the specification in two ways: print all the "jiong" words together after reading in all the inputs; print each "jiong" word once an integer indicating the nesting level of the corresponding "jiong" word is input.

This typical scenario is prone to trigger the error mode of "Selectivity". For the same task content, pieces of information are scattered at different places. Some pieces of information are redundant or incomplete. Under such circumstances, it is hard for a person to collect all the information, exclude redundant information and form an accurate mental representation. As a result, one may selectively notice only a fraction of the whole body of information which are psychological salient, and thus commit errors. Under such situation, some programmers may just notice the information of "Sample inputs" and "Sample outputs" as they are visually presented, and visual representations are generally more likely to capture people's attention, and thus, wrongly interpret the specification as "print all the 'jiong's together after reading all of the inputs."

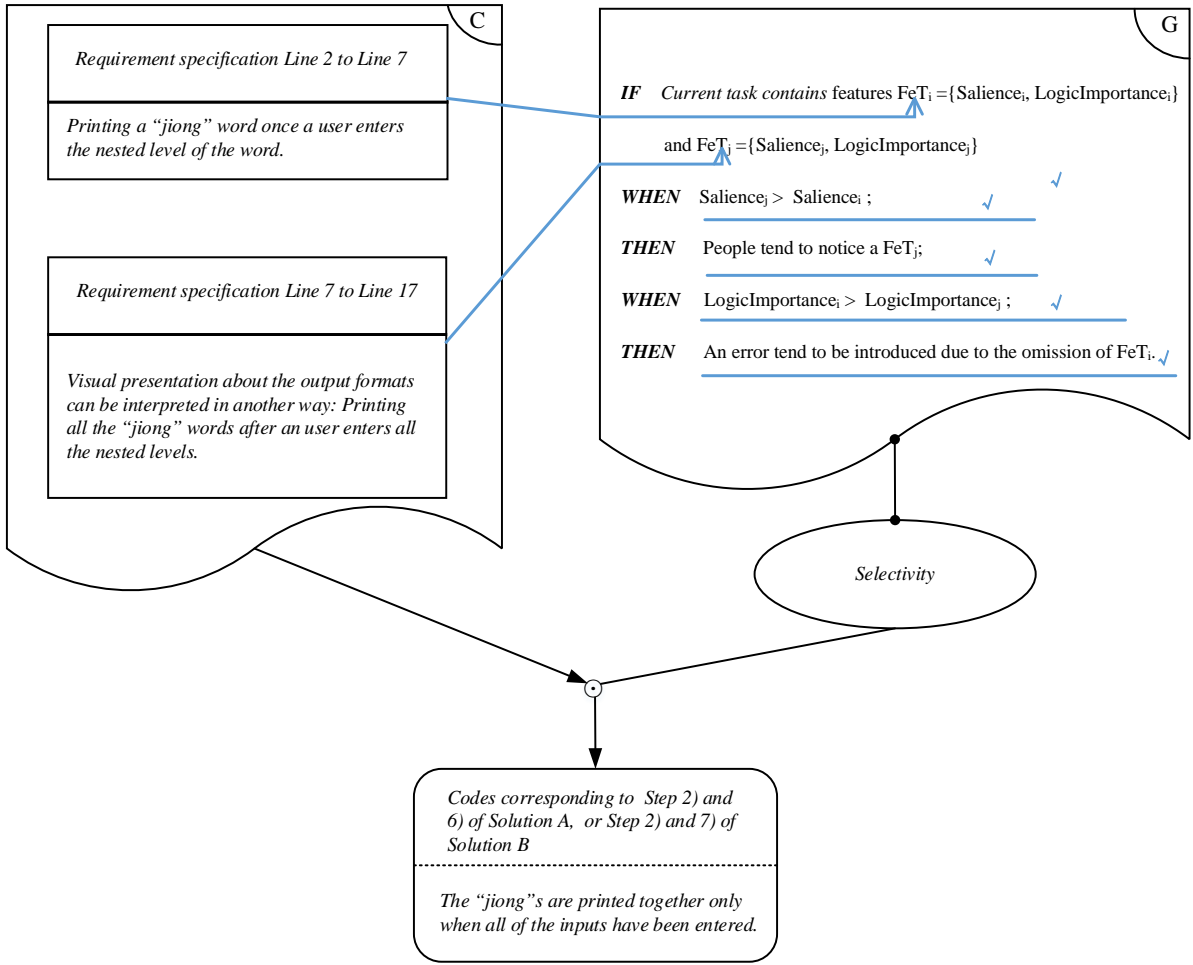

***Defect form:*** *The "jiong"s are printed together only after all of the inputs have been entered.*

TABLE 11
ERROR-PRONE SCENARIO ANALYSIS (7)

*Scenario ID:* ES7

*Defect location:* In the algorithm where it requires modeling the relation between the height of a "jiong" and the "jiong" nesting level, i.e. Step 4) of Solution A or Step 3) of Solution B in Table 2.

*Error modes:* Difficulties with exponential developments and biased review

*Scenario analysis:*

In the "jiong" task, programmers need to model the relationship between the height (h) and nesting level of the "jiong" (n). The right model should be $h=2^{n+2}$. This is an exponential model which must be built if a subject wants to solve the "jiong" problem successfully. This is a typical scenario for the error mode "difficulty with exponential developments" described in Table 1 (Reason, 1990).

Meanwhile, in problem solving, people always evaluate their solutions after constructing an initial solution (F. Huang et al., 2012). When evaluating solutions, humans tend to "biased review" (the error mode described in the third row in Table 1): believing that all possible cases have been checked while in fact very few have been considered.

Notice that there are three groups of sample inputs (nesting level n=1, 2, 3, respectively). The tendency to "biased review" of solution implies that some people may choose a subset of the sample inputs (instead of all the three sample inputs) to check their solutions. But for n=1 or n=2, there is a linear model $h=8n$ that produces the same value as the exponential model $h=2^{n+2}$. Therefore, the wrong model may escape one's self review if one uses only the first two sample inputs to test his solution. Based on these scenarios, the analyst anticipates that some programmers may finally model the relationship between the height and nesting level wrongly as $h=8n$, instead of $h=2^{n+2}$.

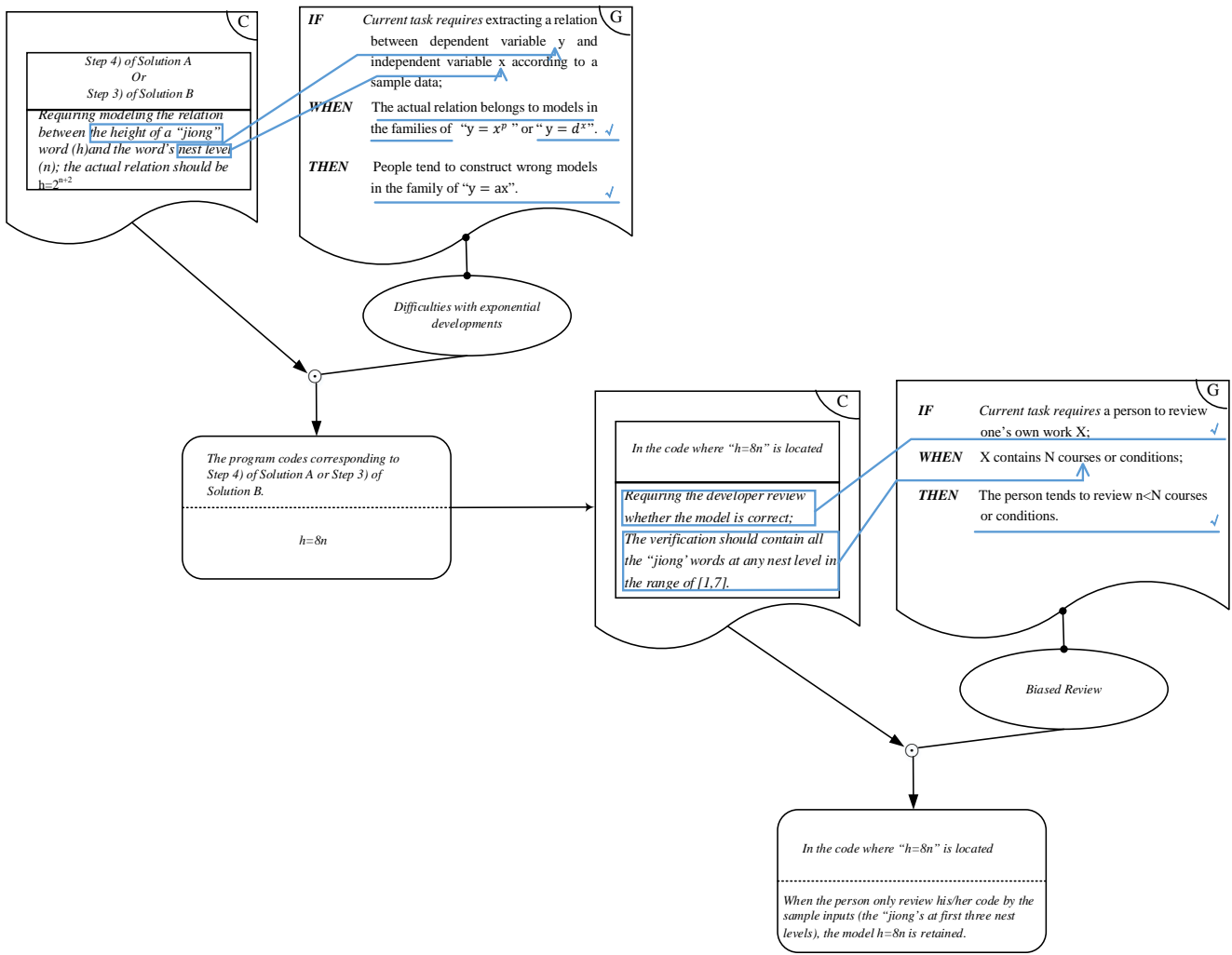

*Defect form:* The relationship between the height and nest level is modeled wrongly as $h=8n$, instead of $h=2^{n+2}$

## 4.6 Actual defects introduced by the participants

A software engineer independent of this paper performed code inspection to identify the defects introduced by the contestants on the "jiong" task. As the contestants were allowed to debug and re-submit their programs, each person can submit more than one version. The code inspector was asked to review all the versions for a contestant, and recorded all of the defects contained in these versions. This is because we were concerned with the error-proneness of programming activities, hence we were interested in all of the errors a contestant made during the entire process of solving the "jiong" problem. However, for each contestant, each defect was counted only once in the statistics that follow, even if it appeared in several versions.

The defects found in the code inspection are summarized in the first four columns of Table 12.

TABLE 12

THE ACTUAL DEFECT DATA

| Defect ID | Defect Description | #Occurrences | Rate of Occurrences | Was this defect predicted? |
|---|---|---|---|---|
| F1 | The size of an array is smaller than that is required: the array size is defined as 100x100, where 512x512 is needed. | 7 | 12.7% | Yes: ES3 |
| F2 | The blank line after the "jiong" is missing | 23 | 41.2% | Yes: ES1 |
| F3 | Mistaking the symbol "!" for "\|"when printing the "jiong" | 1 | 1.8% | No |
| F4 | The symbol '+'in the last line of "jiong" is missing | 1 | 1.8% | No |
| F5 | A number indicating the height of the word "jiong" was printed out by mistake. | 1 | 1.8% | No |
| F6 | Array or variable is used without being initialized. | 14 | 25.5% | Yes: ES2 |
| F7 | Array is initialized by the wrong value, by "0" instead of " "(blank space). | 3 | 5.5% | Yes: ES4 |
| F8 | A variable "n" is used as the upper limit in the "for" loop, without initialization. The "n" may have a large value, which causes the iteration time to exceed the allowed ceiling. | 1 | 1.8% | No |
| F9 | The array used to store the "jiong" is initialized by a 2-depth loop, leading to a problem that the program execution time exceeding the time limit of 1000ms. | 2 | 3.6% | Yes (but not necessarily HEDP) |
| F10 | The function "pow ()" (a C Standard Library function computing the power of a number) is used many times in "for" loop and they are called in every iteration, resulting the program running time exceeding the allowed time limit. | 2 | 3.6% | No |
| F11 | Using a counter "n++" for "n--" in "for" loop. | 1 | 1.8% | No |
| F12 | The program is in c++, but the file name is ended ".c". | 1 | 1.8% | No |
| F13 | Array referencing mistake: char c[1000][1000]='0', where the programmer has unintentionally initialized the first value of the array to be "0". | 1 | 1.8% | No |
| F14 | The "memset" function is misused. | 1 | 1.8% | No |
| F15 | Mistaking the "\" for "\\". | 1 | 1.8% | No |
| F16 | "y<=m && m<y+b/2" is expressed as "y<=m<y+b/2" by mistake. | 1 | 1.8% | No |
| F17 | Enumerate the "jiong" words one by one, but the code is incomplete for printing all of the "jiong" words from nesting level 1 to 7. | 2 | 3.6% | Yes: ES5 |
| F18 | Misunderstanding the requirement: printing out all the "jiong"s after all the inputs are entered by the user, while the requirement specifies that the program should print each "jiong" after the nest-level is entered by the user. | 2 | 3.6% | Yes: ES6 |
| F19 | Mistaking the nesting level iteration "n=n/2" for "n=n-1" in "for" loop. | 1 | 1.8% | No |
| F20 | The location for the symbol "\" is wrong, mistaking f4 [32-i][65-i]='\\' for f4[32-i][63-i]='\\' . | 1 | 1.8% | No |
| F21 | The relationship between the height and nest level of the "jiong" is deduced wrongly as h=8n, which is supposed to be $h=2^{n+2}$. | 2 | 3.6% | Yes: ES7 |
| F22 | Slips in indexing array elements, mistaking map[l][1] for map[l][i]. | 1 | 1.8% | No |

## 4.7 Analysis of Results

A total of 55 programmers submitted a total of 192 programs for the "jiong" problems. The code inspector found that 22 defects (the second column in Table 12) constituted *the set of all defects* in the case study.

We call **Occurrence** of a defect (shown in in the third column of Table 12) the number of programmers who introduced that defect in at least one version of their respective submissions. Any defect for which Occurrence is two or more is called a *Coincident Defect*. Among the 22 defects, 9 were *Coincident Defects*.

The *Rate of Occurrence* of a defect (in the fourth column of Table 12) is the percentage of participants who introduced the corresponding defect, defined in (1):

$$\text{Rate of Occurence} = \frac{Occurence}{Total\ Paticipants} \times 100\% \qquad (1)$$

For instance, defect F2 "the blank line after the 'jiong' is missing" was introduced by 23 programmers, that is, 42.1% of the total 55 contestants. The sum of the *Occurrences* for *the set of all defects* is 70.

The actual defects introduced by the programmers were then compared with those predicted through EPSA, shown in the fifth column of Table 12.

The effectiveness of HEDP is then evaluated from five viewpoints as follows.

### 4.7.1 How many unique defects can HEDP predict among all the potential defects?

We wish to know, given the set of all defects "possible" in this piece of software, what fraction is covered by the prediction effort. This is estimated by the percentage of defects predicted by HEDP out of the total actual defects occurred, shown in (1).

$$\text{Coverage of Unique Defects} = \frac{\text{Number of predicted defects}}{\text{Number of actual defects}} \times 100\% \qquad (2)$$

The results of the Coverage of Unique Defects is summarized in Table 13.

TABLE 13

THE RESULTS OF COVERAGE OF UNIQUE DEFECTS

| Predicted defects | Unique defects actually present | | Coverage of Unique Defects | |
|---|---|---|---|---|
| | Coincident defects | Total defects | For predicting coincident defects | For predicting all defects |
| 7 | 9 | 22 | 77.8% [a] | 31.8% [a] |

Among the 22 total defects, 7 defects (31.8% of the total defects) have been predicted by HEDP. We found impressive that by using information about human error mechanisms we were able to predict the accurate locations and forms of these seven defects just on the basis of just the requirement specification and design analysis. This can hardly be achieved by other approaches such as program metric-based models.

Interestingly, each of the seven predicted defects is a "coincident defect"--introduced by more than two programmers in the same way. These seven defects constitute 77.8% of the total coincident defects. This suggests that HEDP is especially effective in predicting those common defects that tend to be introduced by multiple people.

We also note that the set of all *possible* defects for a normal software development task is probably very large. Our small "jiong" problem had 22 defects manifested, while the *possible* defects for an average size industrial task could be many more. Predicting them all could suggest very expensive countermeasures with little extra benefit in any one specific project; predicting those defects that tend to be highly likely to occur seems to be appropriate for cost-effective countermeasures. We will try to capture this quality of predicting defects that are frequent in the next measure of effectiveness.

In this study, all predicted defects did occur in one or more programs. That is, the only prediction errors were "false negatives": defects that had not been predicted occurred. In general, we should expect that some predicted defects may not occur: "false positives". "False positives" did not occur in our study, insofar as all predicted defects appeared at least once. A more important concern is how many false positives should be expected in a specific development of one program only. We will return to this later.

### 4.7.2 How many defect occurrences can HEDP predict among all the potential occurrences?

This is estimated by the percentage of the occurrences of predicted defects out of the Total Occurrences of the actual defects,

shown in (3).

$$\text{Coverage of Defect Occurrences} = \frac{\text{Occurences of predicted defects}}{\text{Occurences of the actual defects}} \times 100\% \quad (3)$$

The results of CDO are summarized in Table 14.

TABLE 14
THE RESULTS OF COVERAGE OF DEFECT OCCURRENCES

| Occurrence of the predicted defects | Occurrences of the actual defects | | *Coverage of Defect Occurrences* | |
|---|---|---|---|---|
| | Coincident defects | Total defects | Occurrence prediction for coincident defects | Occurrence prediction for total defects |
| 53 | 57 | 70 | 93.0% | 75.7% |

This measure is of interest as a rough estimate of the expected fraction of defects that this method can predict in the development of a single program (in a context similar to this case study, of course). An alternative estimate is given later as "*Coverage per Programmer*".

An interesting finding is that the faults predicted by HEDP tend to occur frequently in our sample of programmers. Though the predicted defects constitute only 31.8% of the total actual defects, their total Occurrences constitute 75.7% of the total Occurrences for all the defects, and 93.0% of the Occurrences of coincident defects. This suggests that HEDP was especially effective in predicting those software defects that were most likely to occur. So, compared to *Coverage of Unique Defects*, the *Coverage of Defect Occurrences* suggests how common the predicted defects are, that is, by implication, how useful it is to take measures to prevent them, both for standard software development (for which *Coverage per Programmer* below is the main measure of interest) and for multiple-version development, as used for some critical applications (Strigini, 2005), where avoiding common defects is the main concern.

*4.7.3 How difficult are the predicted defects for the programmers to debug, once the defects are introduced into programs?*

This is estimated by the Average Persistence of Predicted Defects (APPD)—their tendency to remain through successive versions of a program once they have occurred, shown in in (4).

$$\text{APPD} = \frac{\sum_{i=1}^{7} DP_i}{7} \quad (4)$$

where $DP_i$ is the Degree of Persistence (DP) of the defect $i$, and 7 is the total number of defects forecasted by HEDP. $DP_i$ is in turn estimated by (5).

$$DP_i = \sum_{n=1}^{N_i} \frac{VR_{n,i}}{V_{n,i}} / N_i \quad (5)$$

where $V_{n,i}$ is the total number of versions submitted by the programmer $n$ who introduced the defect $i$; $VR_{n,i}$ is the number of these versions in which the defect $i$ is still present; $N_i$ is the total number of programmers who introduced the defect $i$. The fraction $\frac{VR_{n,i}}{V_{n,i}}$ describes the extent to which the defect $i$ tended to remain in the versions generated by the programmer $n$, named the **Persistence** of defect $i$ for programmer $n$ ($P_{n,i}$).

Note that in this study no defect was introduced during the debugging process, so this simple measure captures well the difficulty of correcting a defect[2]. If programmers do introduce defects during debugging, the measure of persistence should be calculated starting from the version in which they were introduced (e.g. if a programmer produces 6 versions, introduces a defect in the 3rd versions and then removes it in producing the 5th, persistence would be 2/4).

---

[2] As mentioned earlier, the Online Judge System gave very limited information on how a program failed, and we have not analyzed whether the changes made by a programmer for each new version submitted were actual corrections and whether they addressed the actual cause of the previous test failure. In most software development environments, programmers would instead learn at least what "wrong answer" they had produced. In this regard, these numbers are intriguing but certainly not conclusive evidence.

The descriptive statistics of the *Persistence* for all the defects are summarized in Table 15.

TABLE 15
DESCRIPTIVE STATISTICS FOR THE PERSISTENCE OF ALL THE DEFECTS

| Df. ID | Occurrence[1] | Total versions in which the defect is present | Persistence of the defect ($P_i$) | | |
|---|---|---|---|---|---|
| | | | Min. | Max. | Mean ($DP_i$) |
| F1 | 7 | 33 | 0.17 | 0.67 | 0.50 |
| F2 | 23 | 98 | 0.17 | 1 | 0.63 |
| F3 | 1 | 1 | 0.09 | 0.09 | 0.09 |
| F4 | 1 | 1 | 0.25 | 0.25 | 0.25 |
| F5 | 1 | 3 | 0.75 | 0.75 | 0.75 |
| F6 | 14 | 73 | 0.33 | 1.00 | 0.83 |
| F7 | 3 | 36 | 0.25 | 1.00 | 0.75 |
| F8 | 1 | 1 | 0.17 | 0.17 | 0.17 |
| F9 | 2 | 5 | 0.17 | 0.22 | 0.19 |
| F10 | 2 | 2 | 0.50 | 1.00 | 0.75 |
| F11 | 1 | 1 | 0.50 | 0.50 | 0.50 |
| F12 | 1 | 18 | 0.72 | 0.72 | 0.72 |
| F13 | 1 | 4 | 1.00 | 1.00 | 1.00 |
| F14 | 1 | 2 | 0.67 | 0.67 | 0.67 |
| F15 | 1 | 1 | 0.17 | 0.17 | 0.17 |
| F16 | 1 | 1 | 0.17 | 0.17 | 0.17 |
| F17 | 2 | 2 | 1.00 | 1.00 | 1.00 |
| F18 | 2 | 23 | 0.94 | 1.00 | 0.97 |
| F19 | 1 | 1 | 0.50 | 0.50 | 0.5 |
| F20 | 1 | 1 | 0.33 | 0.33 | 0.33 |
| F21 | 2 | 27 | 0.76 | 1.00 | 0.88 |
| F22 | 1 | 1 | 0.50 | 0.50 | 0.50 |

The results for the Average Persistence of Predicted Defects are shown in Table 16 and Figure 1. Among the predicted defects, F1 is the easiest for debugging (DP=0.50) and yet persists through a half of programmers' debugging process, while F17 is the most difficult for debugging and remains until the final versions (DP=1.00). Overall, the *Average Persistence of Predicted Defects* is 0.79, which means these predicted defects were highly persistent through the debugging process, contrasting to the average persistence 0.45 for the non-predicted defects. The programmers took 292 iterations to remove the predicted defects, among a total of 335 iterations to remove all the 22 defects. This suggests that our predicted defects tended to be more difficult for programmers to debug than other defects, as shown in Fig. 4.

TABLE 16
PERSISTENCE OF PREDICTED DEFECTS

| HEDP Predicted defects | Degree of Persistence | Average Persistence of Predicted Defects | Average persistence of non-predicted defects |
|---|---|---|---|
| F1 | 0.50 | **0.79** | 0.45 |
| F2 | 0.63 | | |
| F6 | 0.83 | | |
| F7 | 0.75 | | |

| | |
|---|---|
| F17 | 1.00 |
| F18 | 0.97 |
| F21 | 0.88 |

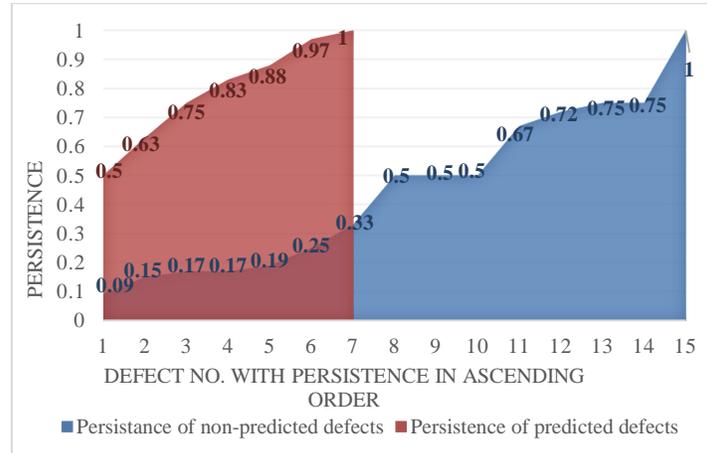

Fig. 4. The persistence of predicted defects compared to that of non-predicted defects

*4.7.4 How many defects can HEDP predict for one programmer?*

This is estimated by the average in (6):

$$Coverage\ per\ Programmer = \frac{\sum_{i=1}^{P_d} \frac{N_{Pi}}{N_{Ai}}}{P_d} \times 100\% \qquad (6)$$

where $N_{Pi}$ is the number of predicted defects for programmer $i$, while $N_{Ai}$ is the number of actual defects introduced by programmer $i$, and $P_d$ the number of programmers who introduced one defect or more. There were 18 programmers who introduced no defect, not relevant for this measure. There were 37 programmers who introduced one defect or more (see Table 17), with a maximum of 5 defects by a programmer, while the median is 1.

Among the programmers who introduced any defects, the percentage of predicted defects for a programmer ranges from 0% to 100%, with mode 100%. The 0%s occurred mostly for programmers who introduced only one defect. For the two worst programmers who introduced 5 defects, HEDP predicted 60% and 80% of those defects. Averaging between programmers, HEDP predicted on average 75% of the defects introduced by a programmer.

The frequency of *false positives* is also important (86%, averaged over the programmers), since they would encourage effort to be spent in preventing errors that did not occur. We must of course consider that occurrence of defects is a random process (involving which programmer happens to be assigned that programming task, plus that programmer's state of alertness, distraction, fatigue, etc., when performing it, plus randomness in the mental processes themselves): the purpose of taking precautions is to avoid errors that *could* happen with high enough probability to be a concern. So, we would expect that this extra effort would not be minded if tolerable, especially for critical software. Similarly, in software testing it is accepted that most test cases will not reveal a defect. Acceptable trade-offs would have to be recognized through accumulated experience, as with any other method.

TABLE 17
THE DISTRIBUTION OF PREDICTED DEFECTS ACROSS ACTUAL DEFECTS

| N. of Prgr[1] | Pst. of D-predi. for a prgr[2] | Detailed observations | | | | Summary measures | |
|---|---|---|---|---|---|---|---|
| | | D prest. | D prest. & predi. | D predi. | FP[3] | FDR[4] | FNR[5] |
| 18 | N/A | 0 | 0 | 7 | 7 | 100% | 0% |
| 13 | 100% | 1 | 1 | 7 | 6 | 85.71% | 0% |
| 8 | 100% | 2 | 2 | 7 | 5 | 71.43% | 0% |
| 2 | 100% | 3 | 3 | 7 | 4 | 57.14% | 0% |
| 2 | 80% | 5 | 4 | 7 | 3 | 42.86% | 20% |
| 3 | 67% | 3 | 2 | 7 | 5 | 71.43% | 33% |
| 1 | 60% | 5 | 3 | 7 | 4 | 57.14% | 40% |
| 1 | 50% | 2 | 1 | 7 | 6 | 85.71% | 50% |
| 6 | 0% | 1 | 0 | 7 | 7 | 100.00% | 100% |
| 1 | 0% | 2 | 0 | 7 | 7 | 100.00% | 100% |
| Avg. | **75%**[6] | 1.25 | 0.96 | 7.00 | 6.04 | 86% | 17% |

[1]Number of programmers in category: a set of programmers with the same number of defects present and present-and-predicted. [2]Percentage of defects predicted for a programmer. [3]False Positive. [4]False Discovery Rate, defined as fraction of the alarms raised that are false alarms (false positives / (false positives +true positives)). The other typical measure of false positives, "false positive rate", defined as (false positives / (false positives + true negatives) cannot be assessed because the number of "non-errors" in a program would be hard to define. [5]False Negative Rate, defined as (false negatives / (false negatives +true pos-itives)). [6]Coverage per Programmer

*4.7.5 How much debugging effort can HEDP save for one programmer?*

This is estimated by the Average Saving of Debugging Effort (ASDE) in (7):

$$ASDE = \sum_n SDE_n/N \quad (7)$$

where $SDE_n$ is the Saving of Debugging Effort for programmer n, and N is the total number of programmers who introduced one or more defects. $SDE_n$ is in turn estimated by Formula (8):

$$SDE_n = \frac{V'_n}{V_n} \times 100\%$$

$$= \frac{\sum_k V'_{n,k}}{V_n} \times 100\% \quad (8)$$

where $V_n$ is the total number of successive versions submitted by programmer $n$ ; $V'_n$ is the number of versions that have not been needed (the effort saved) if the HEDP prediction results (Table 5-Table 11) had been provided and used to avoid the defect. $V'_{n,k}$ is the number of iterations that directly precede the fixing of a predicted defect $k$. We counted $V'_n$ in a conservative manner: when one or more predicted defects are fixed concurrently with an unpredicted defect in a version, all those iterations that directly precede this version are not counted in the estimate of saving, $V'_n$. In the special case where a programmer submitted only 1 version with defects that was "rejected" by the Online Judge System, we assume the programmer's debugging effort are all contained in that version: 1) if the defects in this version are all predicted, we assume the Saving of Debugging Effort is 1; 2) if this version contains any unpredicted defect, we assume the debugging saving is 0.

For instance, a programmer submitted a total of 6 versions, and had his final version "accepted" by the Online Judge System. Our code inspector found that 3 defects (F2, F6, and F15) were present and were removed across these 6 versions. The programmer's debugging history was recorded as the sequence "N F15 **F2** N N **F6**" shown in Fig. 5, where N denotes a version in which no defect is fixed, while a defect ID indexed from Table 12 (e.g. F15 in Fig. 5) indicates this defect is fixed in the corresponding version (e.g. F15 is fixed in the 2nd version). F2 and F6 were predicted defects while F15 was unpredicted (shown in Table 12), therefore, only the 4th and 5th version which directly preceded F6 were counted as saved debugging effort--$V'_n$.

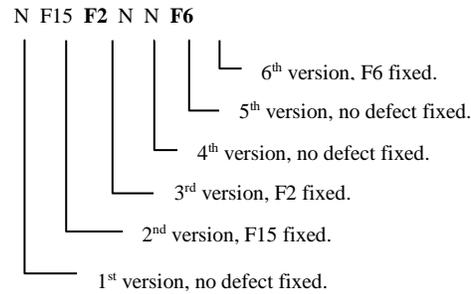

Fig. 5. An example programmer's debugging history

The result lead to an estimate that HEDP would have saved a programmer 46.2% debugging effort on average (Minimum= 0%, Maximum=100%, Standard Deviation=31.4%). Furthermore, HEDP could also significantly increase a programmer's chance of submitting a correct version. Among the total of 37 programmers who committed any errors, 22 programmers got their final version "Accepted" by the Online Judging System, with an Acceptance Rate of 59.5% (22/37). This Acceptance Rate could be increased by 35.1% (13/37), up to 94.6%, if the HEDP were used to avoid or remove predicted defects.

## 5. DISCUSSIONS

### 5.1 Contributions

We return to the two questions proposed at the beginning:

**1). Can the locations and forms of software defects be predicted on the basis of human error mechanisms, before code is produced?**

Yes, in this study. We were able to predict the locations and forms of 75.7% of defect occurrences, and 31.8% of the total defect types, in the "jiong" example. The predicted defects covered, on average over all the programmers who submitted defective programs, 75% of the defects introduced by programmers. Furthermore, these predicted defects were highly persistent through the debugging process: the predicted defects persist through 79% of the debugging process, contrasting to 45% for the non-predicted defects. Most importantly, the prediction was achieved at the requirement phase, using only the information of software requirement specification and the programmers' knowledge background. HEDP could save 46.2% of the debugging effort, using our indicative measure.

The proposed approach demonstrated, in this preliminary study, impressive value in predicting the locations and forms of software defects. We point at two examples. According to standard defect prediction approaches such as program-metric based models, or one's intuitive judgement, "printing a blank line after each 'jiong' word" should be the location where the defects are least likely to be introduced, since it is the simplest and smallest piece of requirements. The fact is, however, that 41.2% participants introduced the same defect at this place, which HEDP predicted to be a high risk location: an instance of a scenario liable to trigger "post-completion error" (Table 5). Another example: based on complexity metrics or intuitive judgment, one may anticipate that the location in a program where the relation between the height of a "jiong" word and its nesting level is dealt with (i.e. Step 4 of Solution A in Table 4) is error-prone, as this is a relatively difficult task point involving mathematical modeling. However, one cannot explain why and how the error takes the form of $h=8n$ (while the correct expression would be $h=2^{n+2}$) rather than any other forms, and how multiple programmers could make the exact same error at this place. Using the Error-Prone Scenario Analysis shown in Table 10, one can clearly predict why and how this location would trigger a defect in this specific form.

In this study, therefore, software defect prediction based on human error mechanisms appears a very promising technique: once the location and form of a software defect can be predicted before the program is produced, the defect can be avoided or prevented in a real sense. Once the error-prone locations are simply flagged to developers (architects, programmers, code reviewers, testers) in advance, they can allocate more attention resources to these locations, thus adjusting the cognitive process to prevent or detect the errors. Another strategy is to improve the representation of the specifications. For instance, at a place where post-completion error is likely to happen, software requirements/specifications can be revised to avoid putting the sub-goal in the last step, or highlighting the sub-goal in the last step to capture developers' attention.

**2).** ***How*** **can software defects be predicted based on human error mechanisms?**

This paper offers one way--HEDP. HEDP predicts software defects through identifying the features in a current programming context that tend to trigger human error modes--cognitive error patterns that psychologists observed to occur in various, diverse activities. This paper extracted an initial set of human error mechanisms, consisting of human error modes and corresponding triggering conditions that are suitable for Error-Prone Scenario Analysis for software defect prediction. We also successfully used them to predict seven defect types in the case study.

We need to emphasize that this paper contributed *one* approach to answer the question, and examined the approach in a case study (which shows the proposed approach workable), but never claim the proposed approach is the sole way. Other approaches may also be effective. For instance, a reader who is a professor teaching a C programming course may say he/she can predict some of the defects in the "jiong" case without the assistance of the proposed approach, e.g. defect F3 in Table 12 "Array is initialized to the wrong value, "0" instead of a blank space". People have some ability to know their own cognition processes (meta-cognition) and learn patterns from their experience; of course, the knowledge gained in such self-learning processes should have some overlaps with what the psychologists learned from studying many people. The fact that a reader can predict F3 without knowing that it is an instance of an error pattern labeled as "Strong-but-now-wrong" by the psychologist Reason (Reason, 1990) neither means that the reader did not implicitly know this erroneous pattern itself through personal experience, nor is relevant to the effectiveness of the proposed approach which can predict F3. A person might predict some of the seven defects based on personal experience, but some seem hard to predict without the systematic theory on human error mechanisms, for instance the post-completion error F2 and exponential development error F 21 in Table 12. HEDP makes it possible for analysts without such "intuitive judgement" or "experience" to predict such defects across diverse project contexts.

This paper *is not intended* to answer the question *"what fraction of the defects can be predicted in a software project?"*. This will vary from case to case. Extrapolating measures of effectiveness from software engineering studies is always an unconvincing exercise. A reasonable claim from this case study is that the HEDP approach has proven effective enough in this case to seem worth more extensive study: via limited trial in an industrial setting, for a software development organization to check whether the approach seems cost-effective in its own environment; and/or via more extensive, controlled studies.

To predict the exact form and location of likely software defects, one needs to look into the causal mechanisms that produce software defects: the characteristics of software requirement (what a programmer is required to do, and how the requirement is presented to the programmer), the knowledge schemata the programmer has, the extent of mismatch between these two factors—the error-prone scenarios produced by the requirements and the knowledge schemata of the specific programmer. That is to say, how many defects can be predicted really depends on the specific features of the software requirement of interest, and the specific features of the programmer who is going to realize the requirement. Using HEDP, we were able to accurately predict 77.8% of defect occurrences for the "jiong" problem performed by a specific group of programmers; this percentage could be higher or lower if the "jiong" problem were solved by another group of programmers with different backgrounds; or on another programming task (requirement), even for the same group of programmers who solved the "jiong" problem. This is not a matter of validity of the proposed approach, but a matter of software's nature.

### 5.2 Cost-effectiveness

An undoubted cost of HEDP is that it requires an analyst with interdisciplinary knowledge of software engineering and of the human error theory used in this study. This may be more than is usually required from a requirement reviewer or code inspector. In the "jiong" case, the analyst had five years' research experience in software psychology and human error theory, as well as eight years' practice in software engineering when she performed the analysis. The extent to which an analyst effectively uses HEDP could affect how many defects he/she can predict for a certain requirement. Future studies should experiment with the training and evaluation of engineers who are potential industrial users.

The overall cost-effectiveness of the approach is determined by the sum of the costs of analysis (which would be determined by the size and complexity of the software specification to be examined) and the cost of the preventative measures adopted, while the benefit would be the cost reductions achieved by the approach. The cost reductions involves 1) the cost of finding and removing those defects that the proactive measures would avoid altogether; or, 2) for defects that are if not avoided, the cost of finding the defects without the benefit of the prediction that helps to focus inspection and testing; or 3) the cost of the defects remaining in the software in operation. This latter cost (of residual defects in the deployed software) varies between the cost of in-operation removal of the defects, plus client aggravation, for much commercial software, and the harm caused by failures in operation, up to potential accidents, with safety-critical software. HEDP does not promise to forecast all potential defects. HEDP is targeted at predicting common defects caused by human cognitive error mechanisms, at early stages of software development, with sufficient accuracy to allow actionable suggestions for prevention or prediction. Defects caused only by random circumstances (e.g., a programmer being interrupted while doing the job and thus writing a line of code at a wrong location) of course cannot be predicted by HEDP, nor any other existing defect prediction methods.

### 5.3 Threats to validity

We discuss four types of validity concerns that apply to by experimental studies in Software Engineering: conclusion validity,

internal validity, external validity and construct validity (Wohlin et al., 2012).

**Conclusion validity** here concerns the degree to which our answers to the two research questions proposed at the beginning of this paper are reasonable. We used an in-depth case study, which is a research methodology especially suitable when one wants to answer "why" and "how" questions and develop novel theories (Runeson & Höst, 2009). We believe the data observed in our study provide sufficient evidence for concluding that: 1) the locations and forms of software defects *can be* predicted on the basis of human error mechanisms before code is produced; and 2) this kind of prediction ***can be*** achieved by our proposed method HEDP. No threat to the validity of these two conclusions is aware of by the authors.

**Internal validity** concerns the extent to which other unknown factors, besides the investigated factors, affect the outcomes. One threat to internal validity is that many factors could influence the likelihood of a programmer committing errors, e.g., interruption, fatigue and time pressure. Such real-time factors are difficult to anticipate before the programmer writing the code, and can hardly link to specific forms of defects that allow for actionable prevention strategies. Defects caused solely by such random circumstances (e.g., a programmer being interrupted while doing the job and writing a line of code at a wrong location) of course cannot be predicted by HEDP nor any other existing defect prediction methods. HEDP does not promise to forecast all potential defects, but is targeted at predicting common defects caused by human cognitive error mechanisms, at early stages of software development, with sufficient accuracy to allow actionable suggestions for prevention. Another threat to internal validity could be that a same defect introduced by different programmers is not caused by a same underlying error mechanism, but because of programmers copying codes from each other or from Internet sources. This threat was avoided by strict anti-cheating procedures (described in Section 4.2).

**Construct validity** concerns the degree to which the specific variables of an empirical study represent the intended constructs in the conceptual or theoretical model. The authors are aware of no threat to the construct validity.

**External validity** concerns the extent to which it is possible to generalize findings and to what extent the findings are of interest to people outside the investigated case. We report results of a single case study, because the results seem impressive enough to justify practitioners experimenting with the approach. In this case study, the predicted defects tended to be "common" defects, introduced by more than one programmer; while defects not predicted were introduced by few, and often by programmers who only introduced few defects. That is, if the results of this study were representative of general effectiveness, the conclusion would be that HEDP was indeed quite effective at predicting (and thus potentially avoiding) defects. How the results would change with e.g. more experienced programmers is a matter of conjecture. We would probably expect all kinds of errors to become much less common; and some kinds almost to disappear. Possibly, experience would teach programmers to recognize themselves many of these error-causing scenarios and thus avoid those errors. However, it is known that common error patterns like post-completion errors are affected by factors like time pressure, common in many development environments, so it would not be surprising if better prevention of even such simple error patterns proved valuable in an industrial context.

### 5.4 Implications

Understanding the human error mechanisms underlying software defects has implications for aspects of defense against software defects. The theory of human error mechanisms constructed and trialed in this paper has implications for three other fields worthy of application.

#### 1). Advancing software defect prevention methods.

The human error mechanisms provided in this paper can be used to improve root cause analysis for defect prevention. Root Cause Analysis (a key process activity in CMMI Level 5) is considered a process to prevent defects. The main processes of this approach include selecting defect samples from historical databases, holding causal analysis meetings to identify the causes of defects, and producing defect prevention strategies for the later projects. The current root cause activities are generally based on root cause taxonomies (F. Q. Huang, Lin, & Huang, 2012). Such taxonomies can only roughly explain *what* factors may have influenced, or related, to the occurrence of software faults, but they are far from explaining the mechanism underlying software defects. Human error mechanisms offer a theory to retrospectively explain *why* and *how* software faults were introduced. Understanding why and how software defects are introduced is certainly important for devising strategies to prevent them. For instance, with the information provided by our Error-Prone Scenario Analysis ES6, one can simply add one notice in Defect Prevention Strategy List (the output document of Root Cause Analysis) like *"**Correct:** printing a "jiong" word once a user enters the nesting level of the word. **Wrong:** the "jiong"s are printed together only after all of the inputs have been entered."* Such a message is highly actionable and would instantly prevent developers from committing this error.

#### 2). Enhancing software testing methods.

Since HEDP produces not only information on the error-prone locations in the program but also possible error forms, they are helpful for test engineers to conduct focused-checks and design test cases. For instance, with the information provided by our error-prone scenario analysis ES7, a code inspector can pay special attention to check whether the mathematical relation between the height of a "jiong" word (h) and its nesting level (n) is $h=2^{n+2}$. That is, the analysis would highlight that this is a requirement for "exponential" programming and thus subject to specific difficulties. With the information of "biased review" provided by ES7, one should pay special attention to design test cases that cover the nesting levels $n>3$ when he/she perform black-box testing. This extends common heuristics for test case selection, like "test all loops using data that require requiring 0, 1, and the maximum number of, iterations."

    **3).Improving requirement specification methods.** Inappropriate specification of software requirements can cause software developers to develop erroneous mental representations(F. Q. Huang et al., 2012), thus leading to software defects that propagate into subsequent development phases. Understanding how inappropriate requirement specifications trigger human errors should play a significant role in reducing the risk of requirements causing defects. Current methods in requirement engineering focus on text ambiguity or  incorrect, incomplete or inconsistent requirements introduced during the requirement elicitation or writing process (Zowghi & Gervasi, 2003). Our study suggests a promising area for identifying during requirement reviews those parts of requirement that tend to trigger human errors, as well as to develop criteria for recognizing error-prone representations to be avoided during requirement writing. For instance, in our study, the simple requirement "printing a blank line after each word" should, according to the current requirement quality criteria such as correctness, completeness, unambiguity and consistency, contain no features prone to trigger a software development error; but has amazingly triggered 23 out of 55 participants to commit the same error in the same way. Once the error-prone representation is identified, one can use strategies to prevent it from triggering development errors. For instance, the requirement writer could highlight (e.g. using bright colors and/or bold font) the places of post-completion tasks in the requirement documents ("printing a blank line after each word" in the "jiong" case), since visual cues are an effective way to reduce post-completion errors (Chung & Byrne, 2008). Though using styles to facilitate readers' cognitive process is not new in software requirement engineering, the contribution here is to tell the writer exactly what should be highlighted to counteract the readers' error-proneness.

## 5.5 Future studies

We report these results from a limited study because they seem impressive enough both to interest other researchers to extend the study in other contexts, and  to make it a reasonable gamble for industrial companies to try out the approach in small pilot trials (e.g. on a subset of some large projects) and see whether it works well for them. For instance, industrial practitioners could use HEDP results to aid document review, and see whether one can better predict and avoid defects than their usual performance based on expertise and intuition.

    Future research could continue to examine the effects of the proposed approach in various ways. One direction is to replicate the same case study with different programmers and HEDP analysts, to examine: 1) whether the list of error kinds forecast here as sufficient for this programming task (e.g., "post-completion error" in Table 5 and "difficulties with exponential developments" in Table 11) are repeatable; 2) whether the errors predicted in Table 6, Table 7, Table 8, Table 9, Table 10 are repeatable if the programmers have knowledge backgrounds similar to that of programmers in our study; 3) whether some new errors can be predicted by using HEDP and extending its base of human error modes; 4) whether any errors exist that are not predicted by HEDP analysts but can be explained retrospectively by the human error mechanisms proposed in this paper; 5) whether there are errors whose causal mechanisms cannot be explained by our human error mechanisms; 6) how many errors can be predicted by other HEDP analysts.

    Another direction is to conduct similar experimental studies with different software specifications and experienced programmers. HEDP predicts defects based on our model of human error mechanisms, which maps the features of diverse programming tasks and programmers to general cognitive error patterns. Thus, we expect HEDP to work for different software specifications and levels of expertise of programmers, and this conjecture is to be tested. Assessing the schemata of experienced programmers could be more difficult that for our somewhat homogeneous set of novice programmers. Future investigations are necessary to validate these kinds of conjectures and test how difficult is to apply HEDP on experienced programmers.

## 6 CONCLUSION

Accuracy in predicting an event typically depends on the extent to which the causal mechanisms underlying the event are understood. This paper proposes an approach to forecasting software defects through the understanding of human error mechanisms that cause them. Compared to established prediction models, which correlate program or process metrics with the likelihood of a program module containing defects, the proposed HEDP approach emphasizes identifying scenarios that tend to trigger human error modes which psychologists have observed to recur across various activities. While current prediction methods mostly suggest software modules to which special attention should be dedicated (especially in testing, or sometimes in reorganizing design before coding) as especially defect-prone, the proposal here is for much more detailed predictions, of specific forms of defect at specific locations in programs, aimed to proactively prevent software defects, through recommending amendments to specifications, or early detection through focused checks in code inspection or test cases. This case study suggests that it is possible to accurately predict many software defects even at the early phases of software development, using the proposed approach. Various measures of effectiveness and potential effort savings are highly encouraging, recommending more trials of this approach in research and industrial contexts.

## ACKNOWLEDGMENT

We thank Mr. Zongquan Ma for his code inspection work and Dr. You Song for giving us access to the programming contest for the case study.